\documentclass[12pt]{article} 
     \usepackage{amsmath}
  \usepackage{graphicx}
  \usepackage{color}
  \newlength{\abstractwidth}
 \usepackage{amsmath,amssymb}
\usepackage{graphicx}
\usepackage{caption}
\usepackage{lipsum}
\usepackage[utf8]{inputenc}
  \usepackage{mciteplus} 
\usepackage{hyperref}

\usepackage[title]{appendix}
\usepackage{subcaption}
\usepackage{comment}
\usepackage{braket}

\def\Xint#1{\mathchoice
   {\XXint\displaystyle\textstyle{#1}}%
   {\XXint\textstyle\scriptstyle{#1}}%
   {\XXint\scriptstyle\scriptscriptstyle{#1}}%
   {\XXint\scriptscriptstyle\scriptscriptstyle{#1}}%
   \!\int}
\def\XXint#1#2#3{{\setbox0=\hbox{$#1{#2#3}{\int}$}
     \vcenter{\hbox{$#2#3$}}\kern-.5\wd0}}

\def\dashint{\Xint-}

\makeatletter
\def\blfootnote{\xdef\@thefnmark{}\@footnotetext}
\makeatother

  \newcommand{\be}{\begin{equation}}
  \newcommand{\bea}{\begin{eqnarray}}
  \newcommand{\eea}{\end{eqnarray}}
  \newcommand{\beq}{\begin{equation}}
  \newcommand{\ee}{\end{equation}}
  \newcommand{\eeq}{\end{equation}}

  \def\ba{\begin{eqnarray}}
  \def\ea{\end{eqnarray}}

 \def\simleq{\; \raise0.3ex\hbox{$<$\kern-0.75em
      \raise-1.1ex\hbox{$\sim$}}\; }
 \def\simgeq{\; \raise0.3ex\hbox{$>$\kern-0.75em
      \raise-1.1ex\hbox{$\sim$}}\; }

\begin{document}

\begin{titlepage}
 % \rightline{}
  \bigskip

  \bigskip\bigskip

  \bigskip

\begin{center}
%\centerline
{\Large \bf{}}
 \bigskip
%\centerline
{\Large \bf {Entanglement Entropy of Two Coupled SYK Models}} 
~\\
~\\
{\Large \bf {and Eternal Traversable Wormhole}} 
\bigskip
\bigskip
   \bigskip
\bigskip
\end{center}

  \begin{center}

 \bf {Yiming Chen$^{1*}$ and Pengfei Zhang$^{2,3\dagger}$}
  \bigskip \rm
\bigskip
 
 \rm

$^{1}$Jadwin Hall, Princeton University,  Princeton, NJ 08540, USA\\
\bigskip
$^{2}$Institute for Advanced Study, Tsinghua University, Beijing, 100084, China \\
\bigskip
$^{3}$Kavli Institute for Theoretical Physics, University of California, Santa Barbara, CA 93106, USA \\

% \vspace{2cm}
  \end{center}

 \bigskip\bigskip
  \begin{abstract}
In this paper, we study the entanglement entropy between two SYK systems with bilinear coupling. We use the replica trick to calculate the entanglement entropy in the ground state. In parallel, we calculate the entanglement entropy through the Ryu-Takayanagi formula in gravity. For the ground state that is dual to an eternal traversable wormhole in AdS$_2$, the bulk quantum correction to the entanglement entropy is of the same order as the minimal surface area. The ground state of the coupled system is close to a thermofield double state with particular temperature and they have the same entanglement entropy. From the gravity point of view, we explain why the two states have the same entanglement entropy. We also study a case with time-dependent coupling, which involves finding the quantum extremal surface in the bulk.
 \end{abstract}
\bigskip \bigskip \bigskip

 \blfootnote{* yimingc@princeton.edu}
 \blfootnote{$\dagger$ PengfeiZhang.physics@gmail.com}

  \end{titlepage}

   % \starttext \baselineskip=17.63pt \setcounter{footnote}{0}
   \tableofcontents
\newpage
 % \sc

\section{Introduction}

In holographic duality \cite{Maldacena:1997re,Gubser:1998bc,Witten:1998qj}, the entanglement structure in the field theory is intimately connected to the geometry of spacetime. The Ryu-Takayanagi (RT) formula \cite{Ryu:2006bv,Ryu:2006ef}
\begin{equation}
     S_{EE} (A) = \frac{\textrm{Area}(\gamma_{A})}{4G_N} 
\end{equation}
states that the entanglement entropy of a boundary region $A$ can be calculated through the area of bulk minimal surface $\gamma_A$ that is homologous to $A$. This prescription of calculating entanglement entropy is derived from AdS/CFT in \cite{Lewkowycz:2013nqa}. The RT formula is generalized to time-dependent geometry as the Hubeny-Rangamani-Ryu-Takayanagi (HRT) formula \cite{Hubeny:2007xt}, in which the requirement for $\gamma_A$ is replaced by being the extremal surface. The formula can also be extended to next order in $G_N$ expansion, where one needs to consider quantum effects in the bulk  \cite{Faulkner:2013ana,Barrella:2013wja,Jafferis:2015del}:
\begin{equation}\label{quantumrt}
    S_{EE} (A) = \frac{\textrm{Area}(\gamma_{A})}{4G_N} + S_{bulk}.
\end{equation}
In the formula, $S_{bulk}$ is the bulk quantum correction of entanglement entropy between the two sides of the surface $\gamma_A$. The two terms in (\ref{quantumrt}) should be extremized together. The formula was conjectured to be true for all orders in $G_N$ in \cite{Engelhardt:2014gca}, and was then proven under AdS/CFT correspondence in \cite{Dong:2017xht}. 

\begin{comment}
The study of the RT formula has deepened our understanding of holographic duality. For example, it motivates the concept of entanglement wedge and the formulation of “subregion-subregion” duality in terms of quantum error correcting code \cite{Almheiri:2014lwa,Dong:2016eik}.
\end{comment}

The interpretation of the RT formula in AdS$_2$/CFT$_1$ appears more difficult than its higher dimensional siblings. One immediate difference is that CFT$_1$ has no spatial degrees of freedom, thus to apply the RT formula, we need to introduce at least two copies of CFT$_1$. This can be achieved by realizing that global AdS$_2$ spacetime has two asymptotic boundaries. A more crucial difficulty lies in the understanding of AdS$_2$/CFT$_1$ correspondence itself, as the pure gravity in AdS$_2$ spacetime being inconsistent with finite energy excitations above the vacuum \cite{Maldacena:1998uz}. However, nearly-AdS$_2$ spacetime is universal in the sense that it describes the near horizon geometry of higher dimensional near-extremal black holes. From this point of view, the Bekenstein Hawking entropy of near-extremal black holes was interpreted in the context of AdS$_2$/CFT$_1$ correspondence \cite{Strominger:1998yg,NavarroSalas:1999up,Sen:2008vm,Castro:2009jf}, and was related to entanglement entropy via the RT formula in \cite{Azeyanagi:2007bj}. There are also other recent progress in understanding entanglement entropy in AdS$_2$ gravity from different perspectives \cite{Callebaut:2018nlq,Callebaut:2018xfu,Lin:2018xkj,Mertens:2019bvy}.

We would like to understand the holographic entanglement entropy in the context of the Sachdev-Ye-Kitaev (SYK) model \cite{PhysRevLett.70.3339,Kitaev,Maldacena:2016hyu}, which shares a common sector as nearly-AdS$_2$ gravity. The SYK model is a (0+1)-d model involving $N$ random interacting Majorana fermions. In low energy, the model has an emergent reparametrization symmetry, which is both explicitly and spontaneously broken. The soft modes associated with this symmetry breaking can be described by the Schwarzian action \cite{Kitaev,Maldacena:2016hyu,Kitaev:2017awl,Jevicki:2016ito,Jevicki:2016bwu}. The same symmetry breaking pattern also appears in nearly-AdS$_2$ gravity \cite{Almheiri:2014cka,Jensen:2016pah,Maldacena:2016upp,Engelsoy:2016xyb}. The gravitational dynamics of nearly-AdS$_2$ is encoded in the movement of the cut-off boundary of the spacetime, which is also described by the Schwarzian action. 

In this paper we study the entanglement entropy between two copies of SYK models coupled by a bilinear term studied in \cite{Maldacena:2018lmt}. More specifically, we will study the entanglement entropy in the ground state of the coupled system. On the gravity side, the coupling between the two boundaries creates a traversable wormhole \cite{Gao:2016bin,Maldacena:2017axo}. Thus the gravity picture for the ground state of the coupled system is an eternal traversable wormhole with global time isometry. Note in order to maintain such eternal traversable wormhole, we need to introduce order $N$ of bulk fields, which makes the bulk quantum correction important in leading order of $N$. We will study the entanglement entropy both from the SYK side and the gravity side. In \cite{Maldacena:2018lmt,Garcia-Garcia:2019poj}, it was shown that the ground state of the coupled system is close to a thermofield double (TFD) state with a particular temperature set by the coupling. Thus both in the SYK and the gravity calculation, we also study the entanglement entropy in the TFD state as a comparison. 

It should be noted that the entanglement entropy of the SYK model was also studied in \cite{Huang:2017nox,Fu:2016yrv} for eigenstates of a single SYK, and in \cite{Gu:2017njx} for a SYK chain. It was also studied for pure states and thermal states in the SYK model and compared with gravity computation \cite{Kourkoulou:2017zaj,Goel:2018ubv}. 

The paper is organized as follows. In section \ref{sec:review}, we provide some review on the boundary and bulk descriptions and set up the notations. In section \ref{sec:SYKcal}, we discuss the calculation of the entanglement entropy in the ground state and the corresponding thermofield double state directly in the SYK model. In section \ref{sec:gravitycal}, we discuss the corresponding gravity calculation and interpretation. We end with a final discussion.

 \section{Review of the boundary and bulk descriptions}\label{sec:review}

In this section, we review some basics of the boundary and bulk descriptions and notations. One can find detailed discussions in \cite{Maldacena:2018lmt}. 

The SYK model contains $N$ Majorana fermions $\psi^i$ with random interaction \cite{Kitaev,Maldacena:2016hyu}:
\begin{equation}
    H_{\textrm{SYK}} = ( i ) ^ { q / 2 } \sum _ { 1 \leq j _ { 1 } \leq j _ { 2 } \cdots \leq j _ { q } } J _ { j _ { 1 } j _ { 2 } \cdots j _ { q } } \psi ^ { j _ { 1 } } \psi ^ { j _ { 2 } } \cdots \psi ^ { j _ { q } },\quad \quad \left\langle J _ { j _ { 1 } \cdots j _ { q } } ^ { 2 } \right\rangle = \frac { 2 ^ { q - 1 } \mathcal { J } ^ { 2 } ( q - 1 ) ! } { q N ^ { q - 1 } }.
\end{equation}
We consider the coupled SYK system as in \cite{Maldacena:2018lmt}:
\begin{equation}\label{coupledH}
\begin{aligned}
    H & = H_{\textrm{L,SYK}} + H_{\textrm{R,SYK}} + H_{\textrm{int}} , \,\,\,\,    H_{\textrm{int}} = i \mu \sum_{j} \psi_{L}^j \psi_R^j.
\end{aligned}
\end{equation}

For small coupling strength $\mu\ll \mathcal{J}$, the low energy physics of the coupled model is governed by the reparametrization modes $t_l (u)$ and $t_{r}(u)$ of the two systems. The dynamics of the reparametrization modes follow the Schwarzian effective action:
\begin{equation}\label{sykeffective}
S = N \int d u \left\{ - \frac { \alpha _ { S } } { \mathcal { J } } \left( \left\{ \tan \frac { t _ { l } ( u ) } { 2 } , u \right\} + \left\{ \tan \frac { t _ { r } ( u ) } { 2 } , u \right\} \right) + \mu \frac { c _ { \Delta } } { ( 2 \mathcal { J } ) ^ { 2 \Delta } } \left[ \frac { t _ { l } ^ { \prime } ( u ) t _ { r } ^ { \prime } ( u ) } { \cos ^ { 2 } \frac { t _ { l } ( u ) - t _ { r } ( u ) } { 2 } } \right] ^ { \Delta } \right\},
\end{equation}
where $c_{\Delta}$ is a number that only depends on $\Delta$, and $\Delta$ is the scaling dimension of the operator that we used to couple the two sides. For the case in (\ref{coupledH}), $\Delta$ is the scaling dimension of the fermion operator, which is $1/q$. One can tune $\Delta\rightarrow p\Delta$ by choosing other coupling terms as $H_{i n t}=g N^{1-p}\left(i \psi_{L}^{j} \psi_{R}^{j}\right)^{p}$. In this paper, we always work in the region that the Schwarzian description is applicable.

The gravity description of the coupled model is in terms of an eternal traversable wormhole in nearly-AdS$_2$ gravity. The metric of global AdS spacetime is
\begin{equation}
    ds^2 = \frac{-dt^2 + d\sigma^2}{\sin^2\sigma},\quad \sigma \in [0,\pi].
\end{equation}
The nearly-AdS$_2$ gravity can be described by Jackiw-Teitelboim gravity \cite{JACKIW1985343,TEITELBOIM198341}:
\begin{equation}\label{JTaction}
  S=\frac{\phi_{0}}{2}\left[\int R+2 \int_{\mathrm{Bdy}} K\right]+\frac{1}{2}\left[\int \phi(R+2)+2 \phi_{b} \int_{\mathrm{Bdy}} K\right]+S_{\mathrm{matter}}[\chi, g]. 
\end{equation}
We parametrize the boundaries by parameter $u$, and the boundary conditions are
\begin{equation}\label{JTboundarycond}
    ds^2|_{\textrm{Bdy}} = -\frac{du^2}{\epsilon^2},\quad \left.\phi\right|_{\mathrm{Bdy}}=\phi_{b}=\frac{\phi_{r}}{\epsilon},
\end{equation}
with $\epsilon$ taken to zero. The term proportional to $\phi_0$ is topological and does not contribute to the equations of motion. $\phi \ll \phi_0$ is the varying part of the dilaton field, which will enter our main discussion. $\chi$ denote the matter fields in the bulk, which do not couple to dilaton field directly. Note the matter fields are necessary to embed JT gravity into the Hilbert space of two copies of the SYK model \cite{Harlow:2018tqv}. The variation with respect to $\phi$ sets $R=-2$, which means the background metric is that of a rigid AdS$_2$ space. The gravitational dynamics is entirely encoded in the movement of the boundaries on this rigid AdS$_2$ space. 

\begin{figure}[t!]
    \centering
    \includegraphics[width=12cm]{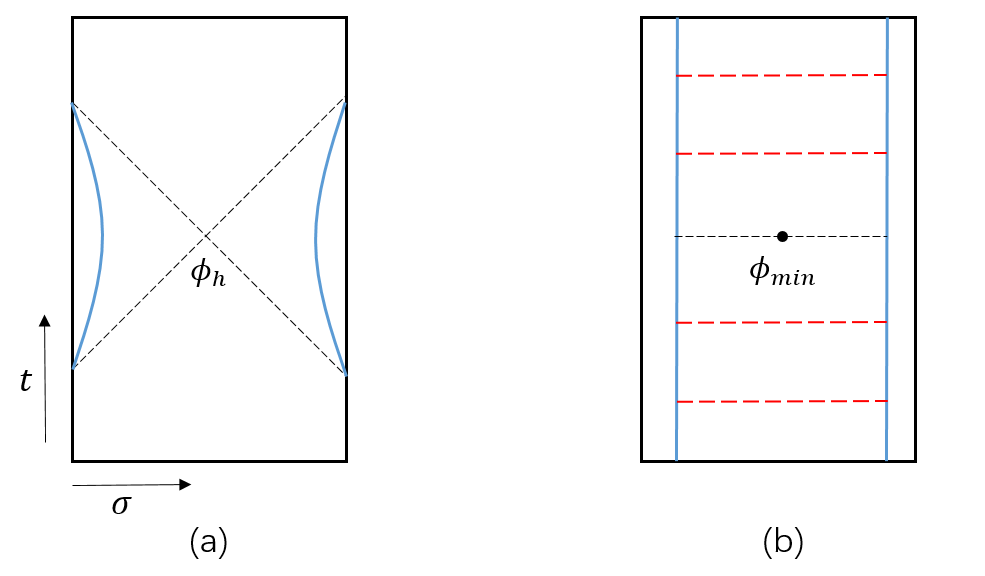}
    \caption{(a) Without boundary couplings, the vacuum solution is a pair of decoupled black holes. The blue lines denote the boundary trajectories on the rigid AdS$_2$ space. For later purpose, we denote the dilaton field on the bifurcation surface as $\phi_h$. (b) The ground state of the coupled model is an eternal traversable wormhole where the boundary trajectories propagate along the global time direction. The red dashed lines denote the couplings. On a constant time slice, the minimal value of the dilaton field lies in the middle of the bulk.}
    \label{fig:gravityintro}
\end{figure}

If the matter fields obey the integrated null energy condition, then we cannot have a traversable wormhole in AdS$_2$ where we can send information between the two boundaries \cite{Maldacena:1998uz}. The vacuum solution of this case is a pair of decoupled black holes (see fig. \ref{fig:gravityintro}(a)). Note here by ``decoupled", we mean that there is no direct coupling between left and right boundaries. The holographic description of this situation is a special entangled state as the thermofield double state \cite{Maldacena:2001kr}.

We can change the situation of non-traversability by adding coupling terms between the two boundaries:
\begin{equation}
    S_{i n t}=g \sum_{i=1}^{N} \int d u\, O_{L}^{i}(u) O_{R}^{i}(u),
\end{equation}
where $O^i$ are a set of $N$ operators with dimension $\Delta$. The effective action with this coupling can be written as:
\begin{equation}\label{graveffective}
    S=\int d u\left\{-\phi_{r}\left\{\tan \frac{t_{l}(u)}{2}, u\right\}-\phi_{r}\left\{\tan \frac{t_{r}(u)}{2}, u\right\}+\frac{g N}{2^{2 \Delta}}\left(\frac{t_{l}^{\prime}(u) t_{r}^{\prime}(u)}{\cos ^{2} \frac{t_{l}(u)-t_{r}(u)}{2}}\right)^{\Delta}\right\},
\end{equation}
where $u$ is the boundary time, or physical time. $t_l (u)$ and $t_r (u)$ are the left/right coordinate times along the boundary trajectories. The ground state of this coupled model is an eternal traversable wormhole where the boundary trajectories propagate along the global time direction.

Following the notation in \cite{Maldacena:2018lmt}, we can rescale the time $u$ and introduce an effective coupling constant $\eta$ to put (\ref{sykeffective}) and (\ref{graveffective}) into the same form:
\begin{equation}
 S = N \int d \tilde { u } \left\{ - \left( \left\{ \tan \frac { t _ { l } ( \tilde { u } ) } { 2 } , \tilde { u } \right\} + \left\{ \tan \frac { t _ { r } ( \tilde { u } ) } { 2 } , \tilde { u } \right\} \right) + \eta \left[ \frac { t _ { l } ^ { \prime } ( \tilde { u } ) t _ { r } ^ { \prime } ( \tilde { u } ) } { \cos ^ { 2 } \frac { t _ { l } ( \tilde { u } ) - t _ { r } ( \tilde { u } ) } { 2 } } \right] ^ { \Delta } \right\},
\end{equation}
with
\begin{equation}
  \tilde{u} \equiv \frac{\mathcal{J}}{\alpha_{S}} u=\frac{N}{\phi_{r}} u, \quad \eta \equiv \frac{\mu \alpha_{S}}{\mathcal{J}} \frac{c_{\Delta}}{\left(2 \alpha_{S}\right)^{2 \Delta}}=\frac{g}{2^{2 \Delta}}\left(\frac{N}{\phi_{r}}\right)^{2 \Delta-1}.
\end{equation}
In the following, we will simply use $u$ to denote this rescaled boundary time. The action should be supplemented by the constraint that the total SL(2,R) charges vanish. The saddle point solution of this action is given by
\begin{equation}\label{globalsol}
    t _ { r } ( u ) = t _ { l } ( u ) = t ^ { \prime } u,\quad\textrm{with}\quad \left( t ^ { \prime } \right) ^ { 2 ( 1 - \Delta ) } = \eta \Delta. 
\end{equation}
Intuitively, the parameter $t'$ quantifies how far away the boundary trajectories are from the boundaries of the rigid space. In this paper, we denote the ground state of coupled system as $\ket{G(\eta)}$. As studied in \cite{Maldacena:2018lmt}, the ground state $\ket{G(\eta)}$ is close to the thermofield double state of two decoupled SYK systems with particular temperature. We denote this thermofield double state by $\ket{TFD(\eta)}$. The temperature of this thermofield double state is related to the coupling by
\begin{equation}\label{tempcoup}
    T(\eta) = \frac{\mathcal{J} }{\alpha_S} \frac{t'}{2\pi} = \frac{\mathcal{J} }{\alpha_S} \frac{ (\eta\Delta)^{\frac{1}{2(1-\Delta)}}  }{2\pi}.
\end{equation}

 \section{SYK calculation}\label{sec:SYKcal}

\subsection{Entanglement entropy of $\ket{TFD(\eta)}$}

For the thermofield double state 
$\ket{TFD(\eta)}$, the reduced density matrix of one side is simply the thermal density matrix. Thus the entanglement entropy between left and right equals to the thermal entropy of the left/right system with temperature $T(\eta)$. By the formula of thermal entropy of a single SYK system \cite{Maldacena:2016hyu}, we have
\begin{equation}\label{tfdee}
   S_{EE} = S _ { 0 } + \frac { N ( 2 \pi ) ^ { 2 } \alpha _ { S } T(\eta)} { \mathcal { J } } = S_0 + 2\pi N (\eta\Delta)^{\frac{1}{2(1-\Delta)}},
\end{equation}
where $S_0$ is the ground state entropy of the SYK model.

\subsection{Entanglement entropy of $\ket{G(\eta)}$}\label{sec:replica}

\begin{figure}[t!]
    \centering
    \includegraphics[width=7cm]{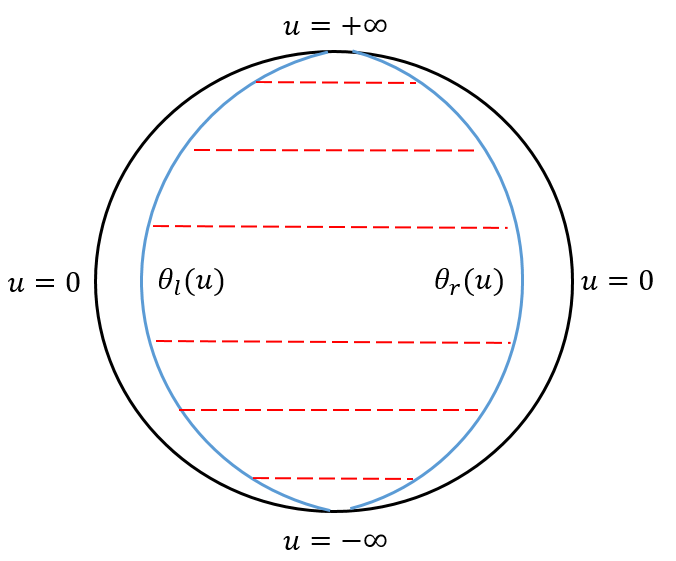}
    \caption{In Euclidean signature, the effective action can be expressed using angular coordinate of a Hyperbolic disk. In the figure, the blue lines represent the two boundary trajectories, and the red dashed lines denote the coupling between them.}
    \label{fig:disk}
\end{figure}

For the state $\ket{G(\eta)}$, we can use the standard replica trick and Euclidean path integral to calculate the entanglement entropy between left and right subsystems. We start from the Schwarzian effective action in Lorentzian signature:
\begin{equation}
 S = N \int d u \left\{ - \left( \left\{ \tan \frac { t _ { l } ( u ) } { 2 } , u \right\} + \left\{ \tan \frac { t _ { r } ( u ) } { 2 } , u \right\} \right) + \eta \left[ \frac { t _ { l } ^ { \prime } (u ) t _ { r } ^ { \prime } ( u ) } { \cos ^ { 2 } \frac { t _ { l } ( u ) - t _ { r } ( u ) } { 2 } } \right] ^ { \Delta } \right\}.
\end{equation}
To use replica method, we transform the above action into Euclidean signature. On the Hyperbolic disk (Euclidean AdS$_2$ space), the Euclidean action is written as:
\begin{equation}\label{diskaction}
    - \frac { S_{E} } { N } = \int _ { - \infty } ^ { \infty } d u \left\{\left( \left\{ \tan \frac { \theta _ { l } ( u ) } { 2 } , u \right\} + \left\{ \tan \frac { \theta _ { r } ( u ) } { 2 } , u \right\} \right) + \eta  \left[ \frac { -\theta _ { l } ^ { \prime } ( u ) \theta _ { r } ^ { \prime } ( u ) } { \sin ^ { 2 } \frac { \theta _ { l } ( u ) - \theta _ { r } ( u ) } { 2 } } \right] ^ { \Delta } \right\},
\end{equation}
where $\theta$ is the angular coordinate of Hyperbolic disk (see fig. \ref{fig:disk} for illustration). There's an extra minus sign in the coupling term because $\theta_l(u)$ and $\theta_r (u)$ propagate in reversed directions on the disk. Note again the Euclidean action should be supplemented by the constraint that the total SL(2,R) charges vanish.

One can add a purely topological term to the above action:
\begin{equation}
    -S_{\textrm{top}} = \frac{S_0}{2\pi} \int du\, \theta'.
\end{equation}
When evaluated on a disk, this always gives us a constant term $S_0$. By the following replica calculation, this term accounts for the ground state entropy term in (\ref{tfdee}). For simplicity, we omit this term in following calculation, and focus on the difference between $S_{EE}$ and $S_0$.

\begin{figure}[t!]
    \centering
    \includegraphics[width=16cm]{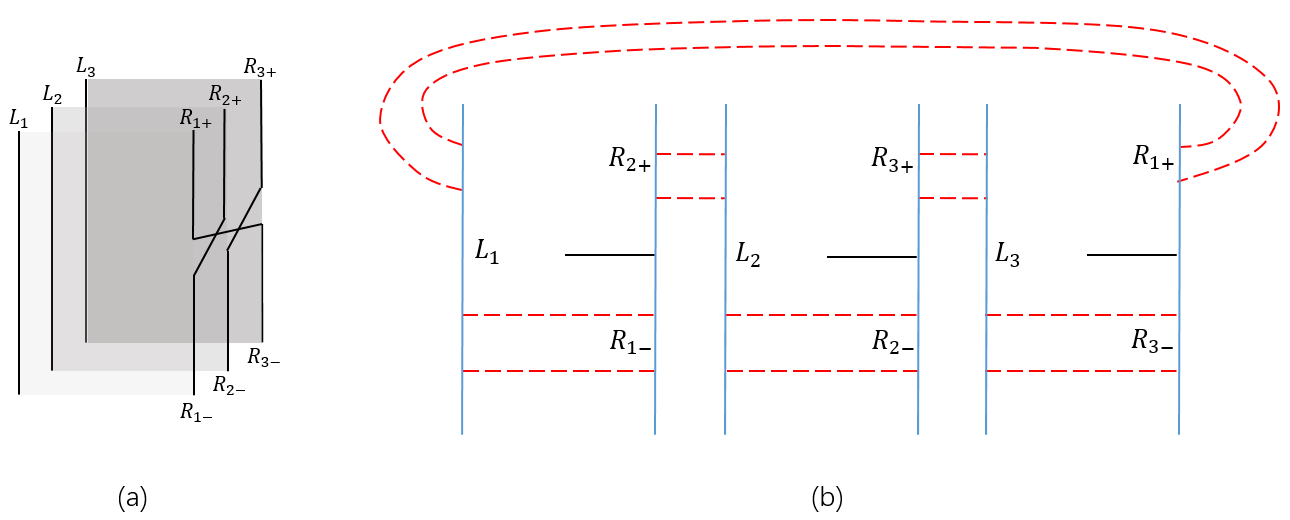}
    \caption{Example with three replicas. In (a), we take three replicas, cut the right boundaries, and identify the boundary condition of $R_{i-}$ with $R_{i+1,+}$. Note the shaded grey area denotes nothing physical. In (b), we draw an equivalent way to picture the configuration. The red dashed lines represent the interaction between boundaries $L_i$ and $R_i$.}
    \label{fig:replicathree}
\end{figure}

To calculate entanglement entropy, we take $n$ replicas of the coupled system. Then we have boundaries $L_i$ and $R_i$, $i=1,...,n$. Note that there is in fact nothing physical in between of $L_i$ and $R_i$, while we still call them as ``boundaries" since their action is the same as the boundary trajectories in rigid AdS$_2$ spacetime. Follow the standard procedure, we cut the right boundaries of each replica at $u=0$, and we denote $R_{i}$ with $u<0$ as $R_{i-}$, $u>0$ as $R_{i+}$. Then we glue $R_{i-}$ to $R_{i+1,+}$ in the path integral ($R_{n+1}$ is identified as $R_{1}$). In fig. \ref{fig:replicathree} (a), we draw the example with three replicas. In fig. \ref{fig:replicathree} (b), we present another equivalent way to picture the configuration.

\begin{figure}[t!]
    \centering
    \includegraphics[width=6cm]{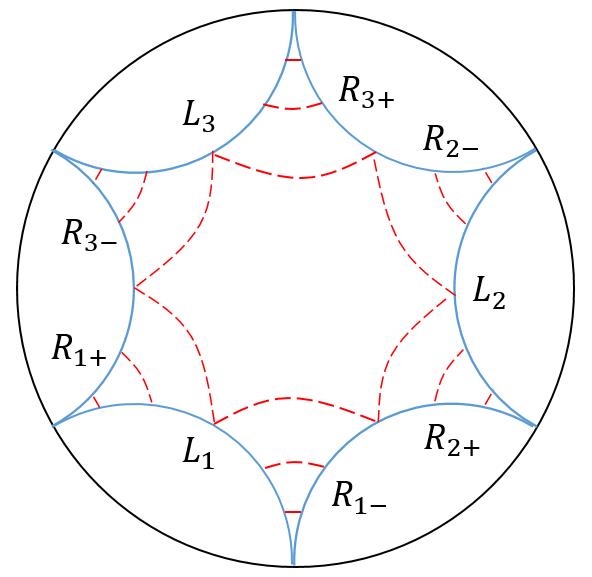}
    \caption{We place the system with three replicas on the Hyperbolic disk. The red dashed lines denote the coupling between the boundaries.}
    \label{fig:threereplicadisk}
\end{figure}

We can also embed the configuration of $n$ replicas in the Hyperbolic disk. In angular order, we place the trajectories $\left\{ L_1 , R_{1-}\cup R_{2+}, L_{2}, R_{2-}\cup R_{3+},... , L_{n}, R_{n-}\cup R_{1+}   \right\}$. In fig. \ref{fig:threereplicadisk}, we show the example with three replicas. After this embedding, we introduce angular functions $\theta_{i} (u)$, $i=1,2,...,2n$ to describe the trajectories. 

The effective action describing the $n$-replica system is
\begin{equation}
      -\frac{S_n}{N} = \int_{-\infty}^{\infty} du \,\sum_{i=1}^{2n} \left\{ \tan \frac{\theta_i (u)}{2},u   \right\}   +  \sum_{i,j}\eta I_{ij}(u) \left[ \frac{-\theta_i '(u) \theta_j '(u) }{\sin^2 \frac{\theta_i (u) - \theta_j (u)}{2}}   \right]^\Delta,
\end{equation}
with the coupling function $I_{ij}(u)$ being
\begin{equation}
    I_{ij} (u) = \left\{
\begin{aligned}
    & H (-u)\delta_{i+1,j}\,\quad i= 2k+1,\\
    & H(u) \delta_{i+1,j}\, \quad i=2k,
\end{aligned}\right.
\end{equation}
where $H(x)$ is the Heaviside step function. One can verify this in the example of three replicas in fig. \ref{fig:threereplicadisk}, by reading how the red dashed lines connect the boundaries.

By assuming the $Z_{2n}$ symmetry on the disk is unbroken, we can express the whole action in terms of a single $\theta $ function,
\begin{equation}
   - S_n /N = 2n \int_{-\infty}^{0} du \, \left(2 \left\{ \tan \frac{\theta (u)}{2},u   \right\}  + \eta \left(  \frac{\theta'(u)^2}{\sin^2 \theta (u)} \right)^{\Delta}  \right).
\end{equation}
In writing down this action, we've fixed the boundary conditions of $\theta(u)$ as $\theta (-\infty) = 0$ and $\theta (0) = \pi/(2n)$.

To evaluate the action $S_n$ for general $n$, we need to solve the equation of motion for $\theta (u)$ subjected to the boundary conditions, which is a nontrivial task. However, since we are only looking for the entanglement entropy, which involves taking the $n\rightarrow 1$ limit, we can use the following trick without really solving $\theta(u)$. 

To illustrate the trick, it is clearer to first do a change of variable: $f(u) = n\theta (u)$, then $f (u)$ will have the boundary condition $f(-\infty) = 0, f(0) = \pi/2$, but with a different action
\begin{equation}\label{actionforf}
   - S_n /N = 2n S(f ) \equiv   2n \int_{-\infty}^{0} du \, \left(2 \left\{ \tan \frac{f (u)}{2n},u   \right\}  + \eta \left(  \frac{f'(u)^2}{n^2\sin^2 f (u)/n} \right)^{\Delta}  \right). 
\end{equation}
In above, we defined the action $S(f)$ without the $2n$ factor. For the special case of $n=1$, we know the classical solution of $S(f)$, which is related to the solution in eqn. (\ref{globalsol}) by a conformal transformation. We denote the solution of $n=1$ as $f_0$ \footnote{The equation of motion for $f$ is a fourth order equation, while we only imposed two boundary conditions for $f$, so it's not immediately clear whether it is the unique solution up to a free parameter $t'$. However, we can prove that there are no other solutions at least in the vicinity of $f_0$, by considering the zero modes of the action for fluctuations around $f_0$. One finds that these zero modes are incompatible with the two boundary conditions.}: 
\begin{equation}\label{zerothsol}
   f_0 (u) = 2 \arctan (\exp (t'u)). 
\end{equation}

\begin{comment}
The equation of motion for $f$ is a fourth order equation, while we only imposed two boundary conditions for $f$, so it's not immediately clear whether it is the unique solution up to a free parameter $t'$. It's hard to find all solutions of the equation of motion, but we can see whether there exist solutions in the vicinity of $f_0$, by considering fluctuations around it. For our convenience, we write the fluctuations as
\begin{equation}\label{fluctuations}
   f (u) =  2\arctan \left[\exp (t' (u + \epsilon(u))\right].
\end{equation}
The boundary conditions of $f$ give constraint on $\epsilon$: $\epsilon (0) =0$, $\lim_{u\rightarrow -\infty}u + \epsilon(u) = - \infty$. Since we have a free parameter in $f$, we can set $\epsilon'(0)=0$ as well. Taking (\ref{fluctuations}) into the action (\ref{actionforf}) of $n=1$, we find the action for $\epsilon (u)$ up to second order:
\begin{equation}
    S(\epsilon)\propto  \int du \left(\epsilon'^2 + \epsilon''^2 \right).
\end{equation}
The zero modes of this action are $1, u, e^{u}, e^{-u}$. However, it is impossible to combine these modes to satisfy the constraints on $\epsilon(u)$. Thus we reach at the conclusion that in the vicinity of $f_0$, there are no other solutions.
\end{comment}

The free parameter $t'$ in (\ref{zerothsol}) is determined by demanding the total SL(2) charge vanish, which leads to the same result as in eqn. (\ref{globalsol}), i.e.
\begin{equation}
    (t')^{2(1-\Delta)} = \eta \Delta.
\end{equation}

Around $n=1$, the solution of $S(f)$ can be expanded as
\begin{equation}
    f = f_0 + (n-1) f_1 + ...,
\end{equation}
and we can also expand the action around $n=1$, i.e.
\begin{equation}
    S(f) = S_0 (f) + \tilde{S} (f) (n-1) + ...
\end{equation}
To evaluate the entanglement entropy, what we need to compute in the end is the difference $S(f) - S_0 (f_0)$ to the first order of $(n-1)$. We have
\begin{equation}
\begin{aligned}
    S(f) - S_{0}(f_0) & = (S_{0}(f) - S_0 (f_0)) + (n-1) \tilde{S} (f) +\mathcal{O}((n-1)^2) \\
    & =  (n-1) \tilde{S} (f_0)+ \mathcal{O}((n-1)^2) ,
\end{aligned}
\end{equation}
where the first term on the first line vanishes because $f_0$ satisfies the equation of motion of $S_0$. Then we see what we need is only $(n-1)\tilde{S} (f_0) $, which can be evaluated without knowing general solution $f(u)$.

Taking the solution $f_0$ in eqn. (\ref{zerothsol}) into the action $S(f)$, and expanding around $n=1$ explicitly, we find
\begin{equation}
\begin{aligned}
    \tilde{S}(f_0 )  & = \int_{-\infty}^{0}du \left[ - \frac{2}{\cosh^2(t' u)}(t')^2  - 2\eta \Delta     \left( 1 + 2  \arctan(e^{t'u}) \sinh(t'u)\right)(t')^{2\Delta} \right]\\
    & = - 2 t' + \eta (2 -\pi) \Delta  (t')^{2\Delta-1} .
\end{aligned}
\end{equation}
Then the entanglement entropy can be evaluated as
\begin{equation}
\begin{aligned}
       S_{EE} - S_0 & = (n\partial_n - 1)S_n |_{n=1} =  -2N \tilde{S} (f_0) \\
       & = 4 Nt' + 2N\eta (\pi -2) \Delta (t')^{2\Delta -1} \\
       & = 2\pi N (\eta \Delta)^{\frac{1}{2(1-\Delta)}}.
\end{aligned}
\end{equation}
Thus
\begin{equation}\label{EESYKground}
     S_{EE} = S_0 + 2\pi N (\eta \Delta)^{\frac{1}{2(1-\Delta)}}.
\end{equation}
The result is the same as what we got in (\ref{tfdee}). This is saying that $\ket{G(\eta)}$ has the same entanglement entropy as $\ket{TFD(\eta)}$, as long as we are in the region that the coupling is small and the Schwarzian description is applicable.

\section{Gravity calculation}\label{sec:gravitycal}

\subsection{Ryu-Takayanagi formula}

In Jackiw-Teitelboim gravity, the prescription of the RT formula is to find the minimal value of the dilaton field in the bulk. This can be seen by noting that $\phi$ measures the area of the transverse sphere from higher dimensional point of view. In the context of the black hole entropy, the Bekenstein Hawking formula \cite{PhysRevD.46.5383,PhysRevD.51.1781} is 
\begin{equation}
    S_{BH} = 2\pi \phi_h.
\end{equation}
In the action (\ref{JTaction}), the constant value $\phi_0$ of the dilaton field contributes to the extremal entropy, which does not change with the coupling between two sides. In boundary theory, it corresponds to the $S_0$ piece. In below, we will not keep track of its contribution and only focus on the deviation of the dilaton field away from constant value $\phi_0$. 

The RT formula is
\begin{equation}
S_{EE}=2\pi \phi + \delta S_{bulk},
\end{equation}
where $\phi$ is the value of the dilaton field at the position that the sum is extremized, $\delta S_{bulk}$ is the quantum correction to the entanglement entropy, which is measured relative to its value in the vacuum state. Since we introduced a large number (of order $N$) of matter fields in the bulk, the term $\delta S_{bulk}$ will be of order $N$ as the term $2\pi \phi$ when we turn on the coupling.

\subsubsection{Two decoupled black holes}

The gravity dual of the thermofield double state $\ket{TFD(\eta)}$ are two decoupled black holes \cite{Maldacena:2001kr}. The bulk matter fields are in the vacuum state, thus we do not need to consider $\delta S_{bulk}$ in this case. The RT surface locates at the bifurcation surface (see fig. \ref{fig:gravityintro}(a)). We denote $\phi_h$ as the value of dilaton field at the bifurcation surface. By the RT proposal, the entanglement entropy between the two sides is
\begin{equation}\label{eetwodecoupled}
     S_{EE} = 2\pi \phi_h.
\end{equation}
This is also the entropy of the left/right black hole with corresponding temperature. To check this explicitly, we note that $\phi_h$ is related to the renormalized boundary value  $\phi_r$ by
\begin{equation}\label{relatephihphir}
    \phi_h = \frac{2\pi \phi_r}{\tilde{\beta}}.
\end{equation}
The rescaled inverse temperature $\tilde{\beta}$ is related to the inverse temperature of the thermofield double state by \cite{Maldacena:2018lmt}:
\begin{equation}
    \tilde{\beta} =\frac{\mathcal{J}}{\alpha_S}\beta.
\end{equation}
Through Schwarzian analysis \cite{Maldacena:2016upp}, the entropy of a single black hole is given by
\begin{equation}\label{entropysinglebh}
    S = \frac {  ( 2 \pi ) ^ { 2 } \phi _ { r }} { \tilde{\beta} }.
\end{equation}
Combine (\ref{entropysinglebh}) and (\ref{relatephihphir}), we get (\ref{eetwodecoupled})
as expected. Using the relation between temperature and coupling in (\ref{tempcoup}), we can express $\phi_h$ in terms of $\eta$:
\begin{equation}\label{phih}
    \phi_h = N (\eta\Delta)^{\frac{1}{2(1-\Delta)}}.
\end{equation}

\subsubsection{Eternal traversable wormhole}\label{sec:rteternal}

The gravity dual of the ground state $\ket{G(\eta)}$ is an eternal traversable wormhole, with global time-translational symmetry. To support the wormhole, we need to introduce a large number (of order $N$) of matter fields in the bulk. By coupling the two sides, we are effectively changing the boundary condition of the fields \cite{Berkooz:2002ug,Witten:2001ua}, and the negative energy in the bulk can be viewed as the decrease of the Casimir energy when we change the boundary condition.

To apply the RT formula, we first need to solve for the dilaton profile in the bulk, which requires knowing the energy distribution of matter fields induced by the coupling. In general, we can do a perturbative calculation in terms of the coupling, which is done for free scalar fields in AdS$_2$ in the appendix. \ref{sec:massive}. Different $\Delta$ corresponds to a different mass of matter field in the bulk. 

In this section, we will look at a special case of $\Delta = 1/2$. In this case, we can choose the matter fields to be massless fermions, and we have a conformal field theory on AdS$_2$. Then we can first do the computation of the energy distribution on a flat strip, and then transform it to get the results for AdS$_2$ by
\begin{equation}
    T_{\mu\nu}^{\textrm{AdS}_2} = T_{\mu\nu}^{strip} + \frac{c}{24\pi} 
    \begin{pmatrix}
    1 & 0 \\
    0 & 1
    \end{pmatrix} - \frac{c}{24\pi}g_{\mu\nu}, 
\end{equation}
where $c$ is the central charge of the conformal field on the strip, $g_{\mu\nu}$ is the AdS$_2$ metric. Note if we look at the piece of stress tensor that is induced by the coupling between two sides, it has the same form on AdS$_2$ as the strip.

This calculation of stress tensor for the special case of $\Delta = 1/2$ with massless fermions is treated in the appendix. C in \cite{Maldacena:2018lmt}. Here we briefly summarize the results. We consider $N$ free real fermions on the flat strip. When we don't have the coupling between the two sides, the normal boundary conditions for each fermion field are:
\begin{equation}
    \psi _ { + } = \left. \psi _ { - } \right| _ { \sigma = 0 } , \quad \psi _ { + } = - \left. \psi _ { - } \right| _ { \sigma = \pi }.
\end{equation}
The coupling between left and right modifies the boundary conditions to
\begin{equation}\label{bdycondition}
    \left. \psi _ { + } \right| _ { \sigma = 0 } = \cos \pi \epsilon \left. \psi _ { - } \right| _ { \sigma = 0 } - \sin \pi \epsilon \left. \psi _ { + } \right| _ { \sigma = \pi } , \quad \left. \psi _ { - } \right| _ { \sigma = \pi } = - \cos \pi \epsilon \left. \psi _ { + } \right| _ { \sigma = \pi } - \sin \pi \epsilon \left. \psi _ { - } \right| _ { \sigma = 0 }.
\end{equation}
The parameter $\epsilon$ can be related to the coupling constant $\eta$ by $\epsilon/4\approx \eta$ for small $\eta$. The stress tensor of the matter fields induced by the coupling between two sides is given by
\begin{equation}\label{energydis}
    T_{tt}^M (\sigma) =T_{\sigma\sigma}^M (\sigma) =  - \frac { N } { 4 \pi } \epsilon ( 1 - \epsilon ),\quad T_{t\sigma}^M =0.
\end{equation}
We add superscript $M$ to stress that it is the stress tensor of matter fields. As commented above, the part of the stress tensor on AdS$_2$ that depends on the coupling has the same form as (\ref{energydis}). Then by using the equation of motion of the dilaton field:
\begin{equation}
      \nabla_{\mu} \nabla_{\nu} \phi - g_{\mu\nu} \nabla^2 \phi +g_{\mu\nu} \phi =  -T_{\mu\nu}^{M},
\end{equation}
one finds the dilaton profile as
\begin{equation}
    \phi = N \frac { \epsilon ( 1 - \epsilon ) } { 4 \pi } \left[ \frac { \left( \frac { \pi } { 2 } - \sigma \right) } { \tan \sigma } + 1 \right].
\end{equation}
By the symmetry between left and right, the RT surface locates at $\sigma = \pi/2$ (see fig. \ref{fig:gravityintro}(b)), which is given by
\begin{equation}
    \phi_{min} =  N \frac { \epsilon ( 1 - \epsilon ) } { 4 \pi } .
\end{equation}
For small coupling $\eta$, we have
\begin{equation}\label{rtground}
    \phi_{min} = N \frac{\eta}{\pi}.
\end{equation}
This is the RT contribution to the entanglement entropy in the ground state of the coupled system. 

As a comparison, in the case of two decoupled black holes in the previous section, if we plug $\Delta = 1/2$ into eqn. (\ref{phih}), we have
\begin{equation}\label{phihdelta1/2}
    \phi_h = N \frac{\eta}{2}.
\end{equation}
$\phi_{min}$ and $\phi_h$ have a difference of order $N$: 
\begin{equation}\label{deltaRT}
    \delta \phi \equiv \phi_{min} - \phi_h =  - N  \left( \frac{1}{2} - \frac{1}{\pi}  \right)\eta.
\end{equation}

\subsection{Quantum correction to the entanglement entropy}\label{sec:quantumcorrection}

Since we have $N$ matter fields in the bulk, their quantum correction to the entanglement entropy between the two sides is of the same order as the RT contribution. The quantum correction is the entanglement entropy of the bulk fields on two sides of the RT surface. More precisely, we are interested in the \emph{change} of the entanglement entropy brought by the coupling, which does not contain UV divergence.

For the special case of $\Delta = 1/2$, there are several ways to calculate the quantum correction.  One way is to utilize that for a free fermion system, one can calculate the reduced density matrix and entanglement entropy through two-point correlation functions \cite{Peschel_2003}. This calculation is presented in the appendix. \ref{sec:directcorrection}, where the modular Hamiltonian is also computed explicitly. Another way is by using the replica method, in a similar manner as in sec. \ref{sec:replica}. Here we will follow a third approach valid for small coupling. We use the entanglement first law \cite{Blanco:2013joa}:
\begin{equation}
    \delta S_{bulk} = \delta \langle K\rangle,
\end{equation}
which says the change of entanglement entropy can be calculated through the change of expectation value of modular Hamiltonian. We compare the state in interest with the vacuum state of bulk matter fields (with no coupling), for which we know the explicit form of the modular Hamiltonian.

There is a subtlety about using entanglement first law in this case. The bulk fields with nonzero coupling $\eta$ have different boundary conditions as $\eta=0$, so their ground states are not in the same Hilbert space, and the entanglement first law is not immediately applicable. To use the entanglement first law, we imagine the following dynamical process. Starting from the vacuum state with $\eta =0$ in the long past, we turn on the coupling adiabatically to a small constant value $\eta$. Then we get the ground state of the coupled system, where the bulk fields have modified boundary conditions. To compare with the vacuum state with no coupling, we need to turn off the coupling before $t=0$, thus the two states have the same boundary condition and live in the same Hilbert space. Then we apply the entanglement first law to find the change of entanglement entropy. The key point is that when we turn off the coupling, we release some extra energy that must be taken into account.

Since the vacuum state with zero coupling is invariant under the transformation generated by the Rindler Killing vector $\xi$, the modular Hamiltonian of the left Rindler Wedge is related to the stress tensor by
\begin{equation}
    K = 2\pi  \int_\Sigma *\left(   \xi \cdot T^M  \right).
\end{equation}
In the formula, $\Sigma$ is a spatial slice extends from the RT surface to the left boundary. More explicitly, we can choose $\Sigma$ to be the $t=0$ slice from $\sigma =0$ to $\sigma = \pi/2$, and on this slice, we have
\begin{equation}
    \xi^t = \cos \sigma, \quad \xi^\sigma =0.
\end{equation}
Then we have
\begin{equation}\label{MH}
   K  = 2\pi \int_{0}^{\frac{\pi}{2}}d\sigma \, \cos \sigma \langle T_{tt}^{M}(\sigma) \rangle,
\end{equation}
which is simply $2\pi$ times the Rindler energy. We need to find the energy distribution in the bulk to evaluate the expectation value of $K$. The negative energy in the bulk after we adiabatically turn on the coupling is already given in eqn. (\ref{energydis}) for the case of $\Delta  = 1/2$, and in the appendix. \ref{sec:massive} for general $\Delta$. However, the process that we turn off the coupling before $t=0$ will generate some extra energy close to the boundary, which remains to be calculated.

In general, we imagine turning off the coupling before $t=0$ through a short time of order $\beta$. We will present a detailed discussion of this general set up in sec. \ref{sec:slow}. For now, we look at the case where we turn off the coupling $\eta$ suddenly at $t=-0^+$. Then we generate some energy localized at the boundary. The final result will be qualitatively the same as turning off the coupling more slowly. For $\Delta = 1/2$ case with massless fermions, the calculation of $T_{tt}^M (\sigma,t)$ with time-dependent coupling can be simplified using conformal methods. In appendix. \ref{sec:energyCFT}, we discuss how to calculate $T_{tt}^M (\sigma,t)$ for a general CFT on a strip with varying coupling.  Here we simply write down the result for $T_{tt}^M (\sigma,t)$ at $t=0$, in terms of the parameter $\epsilon$:
\begin{equation}\label{energyquench}
     \langle  T_{tt}^M  (\sigma,t=0)  \rangle/N  = -\frac{\epsilon}{4\pi} + \frac{\epsilon}{8} \left(\delta (\sigma- 0^+)  + \delta (\sigma-\pi + 0^+)\right).
\end{equation}
The formula says that we have uniform negative energy in the bulk, and some positive energy localized at the left boundary and right boundary.

Taking (\ref{energyquench}) into the entanglement first law, we get
\begin{equation}\label{qc}
\begin{aligned}
    \delta S_{bulk} & = 2 \pi N \int_{0}^{\frac{\pi}{2}}d\sigma \, \cos \sigma \left( - \frac{\epsilon}{4\pi}\right) + 2\pi N\times \frac{\epsilon}{8} = N\left( \frac{\pi}{4} - \frac{1}{2}   \right)\epsilon\\
    & = 2\pi N\left( \frac{1}{2} - \frac{1}{\pi}  \right)\eta.
\end{aligned}
\end{equation}
In appendix. \ref{sec:directcorrection}, using a different method, we also arrive at the same result.

Taking (\ref{rtground}) and (\ref{qc}) into the RT formula, we find
\begin{equation}
    S_{EE} = 2\pi N \frac{\eta}{\pi} +  2\pi N\left( \frac{1}{2} - \frac{1}{\pi}  \right)\eta = \pi N  \eta.
\end{equation}
After adding the ground state entropy $S_0$ coming from the constant piece of the dilaton field, this matches with the result (\ref{EESYKground}) in the SYK calculation (with $\Delta = 1/2$).

As a side comment, note that from eqn. (\ref{energyquench}), the global energy at $t=0$ vanishes to the first order of $\epsilon$:
\begin{equation}
     \delta E_{t}^{bulk} = \int_{0}^{\pi} T_{tt}^{M} (\sigma)\, d\sigma =0 + \mathcal{O}(\epsilon^2).
\end{equation}
This is consistent with the statement that the state we get is close to the vacuum state, since to the linear order of $\epsilon$, the process of turning on and off the coupling does not raise the total energy. From this, one can also see that turning off the coupling is essential to make the comparison of energy meaningful.

\begin{comment}
Note that the vacuum energy of conformal field on a strip is negative due to Casimir effect, while this negative energy is exactly canceled by conformal anomaly when we consider vacuum state of conformal fields on AdS$_2$ (as shown in appendix C.1 in \cite{Maldacena:2018lmt}). 
\end{comment}

\subsection{First law and the relation between $\delta \phi$ and $\delta S_{bulk}$}\label{sec:fisrtlawformula}

In the SYK calculation, we find that the two states $\ket{TFD(\eta)}$ and $\ket{G(\eta)}$ have same entanglement entropy. In gravity, the corresponding statement is 
\begin{equation}
   2\pi \phi_h = 2\pi \phi_{min} + \delta S_{bulk},
\end{equation}
or equivalently
\begin{equation}\label{dphids}
   2\pi \delta \phi + \delta S_{bulk} =0.
\end{equation}
For the case of $\Delta = 1/2$, one can check this explicitly from eqn. (\ref{deltaRT}) and (\ref{qc}). The goal of this section is to derive eqn. (\ref{dphids}) for general bulk contents, by deriving the first law in JT gravity. The first law of JT gravity was also discussed in \cite{Callebaut:2018xfu} for other purposes. For the completeness of argument, we present the detailed derivation here. 

We follow the recipe by Wald and Iyer \cite{Wald:1993nt,Iyer:1994ys}. Consider the Jackiw-Teitelboim Lagrangian:
\begin{equation}\label{JTactionWald}
   \mathbf{L} = \frac{1}{2} \mathbf{\epsilon}\phi (R+2),
\end{equation}
where $\mathbf{\epsilon}$ is the volume element associated with the metric. The first order variation of the Lagrangian can be expressed as
\begin{equation}
 \delta \mathbf{L} = \mathbf{E}_g^{ab} \delta g_{ab} + \mathbf{E}_\phi \delta \phi + d \mathbf{\Theta} (g,\phi,\delta g,\delta\phi),
\end{equation}
where $\mathbf{E}_g$ and $\mathbf{E}_{\phi}$ are the equations of motion of the action (\ref{JTactionWald}), and $\mathbf{\Theta}$ is
\begin{equation}
   \mathbf{\Theta}_a = \frac{1}{2}\mathbf{\epsilon}_{ba} \left[ \phi \nabla_{c} \delta g^{bc} - \left( \nabla_c \phi \right) \delta g^{bc} - \phi g^{cd} \nabla^{b} \delta g_{cd} + \left(\nabla^b \phi \right) g^{cd}\delta g_{cd} \right].
\end{equation}
Diffeomorphism $\xi^a$ on the manifold can be associated to a Noether current:
\begin{equation}
   \mathbf{J}[\xi] = \mathbf{\Theta}\left(  g, \phi , \mathcal{L}_{\xi} g, \mathcal{L}_{\xi} \phi   \right) - \xi \cdot \mathbf{L}.
\end{equation}
We have
\begin{equation}\label{conservedcurrent}
   \mathbf{J}[\xi]_a = \mathbf{\epsilon}_{ba} \nabla_{c} \left[  \phi \nabla^{[c} \xi^{b]} + 2 \xi^{[c}\nabla^{b]}\phi  \right] -  2\mathbf{\epsilon}_{ba} E_{g}^{bc}\xi_{c} .
\end{equation}
In the derivation, we are free to set $R=-2$ since the background metric is not dynamical. Under the action (\ref{JTactionWald}), the second term in (\ref{conservedcurrent}) vanishes on shell. However, when we perturb the system by turning on the matter fields, it is nonzero, and
\begin{equation}
   \mathbf{J}[\xi]_a = \mathbf{\epsilon}_{ba} \nabla_{c} \left[  \phi \nabla^{[c} \xi^{b]} + 2 \xi^{[c}\nabla^{b]}\phi  \right]  +   \mathbf{\epsilon}_{ba} T^{M,bc}\xi_{c} .
\end{equation}
The current is associated with a Noether charge $\mathbf{Q}$:
\begin{equation}
   \mathbf{J}_a = (d \mathbf{Q})_a +   \mathbf{\epsilon}_{ba} T^{M,bc}\xi_{c},
\end{equation}
where the Noether charge is \cite{Iyer:1994ys,PhysRevD.51.1781}
\begin{equation}
 \mathbf{Q} = - \frac{1}{2} \epsilon_{ab} \left(   \phi \nabla^{a} \xi^{b} + 2 \xi^{a}\nabla^{b}\phi    \right).
\end{equation}
We now restrict attention to the case where $\xi$ is the Rindler Killing vector, and the system is two decoupled black holes, with no matter field excitation. For arbitrary variation $\left\{ \delta \phi , \delta T^{M}\right\}$ above this solution, we have
\begin{equation}
    \int_{\partial \Sigma} \left( \delta \mathbf{Q}[\xi] - \xi\cdot \mathbf{\Theta} \right) = \int_{\Sigma} \epsilon_{ac}\delta T^{M,ab}\xi_{b}   +  \int_{bdy} \left( \delta \mathbf{Q}[\xi] - \xi\cdot \mathbf{\Theta} \right),
\end{equation}
where $\Sigma$ is the codimension one slice which extends from the fixed point $\partial \Sigma$ of the Rindler Killing vector to the boundary of spacetime. Since the metric is rigid under variations, we have $\mathbf{\Theta} =0$, and
\begin{equation}
  \int_{\partial \Sigma}  \delta \mathbf{Q}[\xi]  = \int_{\Sigma} \epsilon_{ac}\delta T^{M,ab}\xi_{b}   +  \int_{bdy} \delta \mathbf{Q}[\xi] .
\end{equation}
We choose the direction of the Killing vector to point towards future, and normalize it such that it acts like a boost around the fixed point. Then we can put the formula into explicit form as
\begin{equation}
  \delta \phi|_{\partial \Sigma} + \int_{\Sigma} d\sigma \sqrt{h} \delta T_{\mu\nu}^{M} n^\mu \xi^\nu  =  n_{\mu} \sigma_{\nu}    \left[  \delta \phi \nabla^\nu \xi^{\mu} + (\nabla^\mu \delta\phi)\xi^\nu -(\nabla^\nu \delta\phi)\xi^\mu  \right]|_{bdy},
\end{equation}
where $n^{\mu}$ is the normal vector of the slice $\Sigma$, $\sigma_{\mu}$ is the normal vector pointing outwards on the boundary. Around the black hole solution, this formula is the first law of black hole \cite{Bardeen:1973gs,Wald:1993nt}, with the left hand side being the change of area plus the change of Rindler energy of the matter field, and the right hand side being the change of the total mass of the black hole.

However, since the equations of motion for $\phi$ are \emph{linear}, the formula, in fact, holds for more general variations of the solution (as long as we keep $\phi \ll \phi_0$). In fact, we can write down a formula that holds for \emph{any} solutions (without considering variations):
\begin{equation}\label{covariant}
   \phi|_{\partial \Sigma} + \int_{\Sigma} d\sigma \sqrt{h}  T_{\mu\nu}^{M} n^\mu \xi^\nu  =  n_{\mu} \sigma_{\nu}   \left[   \phi \nabla^\nu \xi^{\mu} + (\nabla^\mu \phi)\xi^\nu -(\nabla^\nu \phi)\xi^\mu  \right]|_{bdy}.
\end{equation} 
For example, we can apply the formula to the eternal traversable wormhole, where there is in fact no horizon. For concrete computation, we choose the slice $\Sigma$ to be the the spatial slice of $t=0$, with $\sigma\in (0,\pi/2]$. We have
\begin{equation}
    n_{\mu} =  \left\{ -\frac{1}{\sin \sigma}, 0 \right\},\,\,\,\,  \sigma_{\mu} =  \left\{ 0 , -\frac{1}{\sin \sigma}  \right\},  
\end{equation}
and
\begin{equation}
    \xi^{\mu}|_{t=0} = \{\cos\sigma , 0 \}, \,\,\,\, \nabla^\sigma \xi^{t}|_{t=0} = -\sin \sigma.
\end{equation}
For the case of decoupled black holes and eternal traversable wormhole that we've discussed, the RT surface locates exactly at the fixed point of Rindler Killing vector, and thus $\phi|_{\partial \Sigma} = \phi_{min}$. However, this is not essential, since if not, we can always first use the SL(2,R) transformation to move $\phi_{min}$ to the center of the bulk, and then apply the formula. Equivalently, we can also choose other killing vector that has fixed point on $\phi_{min}$. Then we get
\begin{equation}\label{phiminT1}
    \phi_{min} + \int_{\Sigma} d\sigma\, \cos \sigma T_{tt}^{M} (\sigma ,t) = \left(\cos \sigma \partial_{\sigma}\phi + \csc \sigma \phi  \right)|_{bdy,t=0}.
\end{equation}
This formula relates the minimal dilaton value with the Rindler energy, up to a boundary term.

We can put the boundary term into a different form. By using the equation of motion:
\begin{equation}
     \cot \sigma \partial_{\sigma}\phi + \csc^2 \sigma \phi + \partial_{t}^2 \phi = -T_{\sigma\sigma}^{M} (\sigma,t) ,
\end{equation}
and under the situation that $\sin \sigma T_{\sigma\sigma}^{M}$ vanishes near the boundary (as shown in appendix. \ref{sec:massive}), we get
\begin{equation}\label{phiminT2}
      \phi_{min} + \int_{\Sigma} d\sigma\, \cos \sigma T_{tt}^{M} (\sigma ,t) =   - \sin \sigma \partial_{t}^2 \phi |_{bdy,t=0} .
\end{equation}
If at $t=0$ we have $\partial_t \phi=0$, i.e., the boundary trajectory is perpendicular to the $t=0$ time slice, then the term $\partial_t^2 \phi$ can be intuitively interpreted as the ``acceleration" of the boundary trajectory towards the boundary of AdS$_2$ spacetime. For example, in the ground state of eternal traversable wormhole, the boundary trajectory doesn't accelerate towards the boundary, and we have $\partial_t^2 \phi =0$. Indeed, for the example of $\Delta = 1/2$ that was worked out in previous sections, we have
\begin{equation}
    \phi_{min} + \int_{\Sigma} d\sigma\, \cos \sigma T_{tt}^{M} (\sigma ,t) =  N \frac { \epsilon ( 1 - \epsilon ) } { 4 \pi } +\int_{0}^{\frac{\pi}{2}} d\sigma \, \cos \sigma \left(- \frac { N } { 4 \pi } \epsilon ( 1 - \epsilon )\right) =0.
\end{equation}
We can also consider the case of two decoupled black holes, where we have $T_{tt}^M =0$ in the bulk. In this case, the boundary trajectory is accelerating towards the boundary. The dilaton profile is
\begin{equation}\label{tfdphi}
    \phi = \phi_h \frac{\cos t}{\sin \sigma},
\end{equation}
 and we see $-\sin\sigma \partial_{t}^2 \phi |_{bdy, t=0}= \phi_h $ holds.

In fact, we can further show that the boundary term in (\ref{phiminT2}) is equal to minus of the SL(2,R) charge of the boundary trajectory associated with the symmetry generated by $\xi$. Written in global coordinates, we have
\begin{equation}\label{bdysl2r}
  -\sin \sigma \partial_{t}^2 \phi |_{bdy,t=0}  = - Q^{L}[\xi] |_{t=0}=  - \phi_r \frac{-t_l''^2 + t_l't_l'''}{t_l'^3},
\end{equation}
where $\phi_r$ is defined as in (\ref{JTboundarycond}). One can derive this formula by noting that $\partial_u \phi =0$ and $\partial_u^2 \phi =0$, and translate the derivative with respect to $u$ into global coordinates. Since the Rindler energy in (\ref{phiminT2}) is the SL(2,R) charge of matter field $q_{M}^{L}[\xi]$ in the left Rindler wedge, we can reinterpret (\ref{phiminT2}) as
\begin{equation}
   Q_{f}^L [\xi]  = q_{M}^L [\xi] + Q^L [\xi],
\end{equation} 
where $-Q_f^{L} [\xi]$ is the value of dilaton field at the fixed point of $\xi$. This formula is then natural as the total SL(2,R) charge of the bulk should vanish, i.e. $Q^L + Q^R + q_M =0$ \cite{Maldacena:2016upp}. When we only look at the left wedge, we can think of the SL(2,R) charge on the right wedge as living on the fixed point, i.e. $q_{M}^R + Q^R = -Q_{f}^L$. What this formula tells us is that $-Q_{f}^L$ has a simple expression as the value of the dilaton field on the fixed point.

Now we return to the discussion of the relation between $\delta \phi$ and $\delta S_{bulk}$. As we discussed, to compute $\phi_{min}$ and $\delta S_{bulk}$, we first turn on the coupling adiabatically, which creates the traversable wormhole. We turn off the coupling right before $t=0$, then the boundary trajectory will accelerate towards the boundary of spacetime, which makes the wormhole non-traversable again. We expect that for $t>0$, although the dilaton profile in the bulk can be complicated, near the boundary it should have the form similar to (\ref{tfdphi}), thus the boundary trajectory moves as that of two decoupled black holes with the corresponding temperature. This is shown explicitly by Schwarzian calculation in section. 4.3 in \cite{Maldacena:2018lmt}.
In section. \ref{evolution}, we will also check this expectation by solving the evolution in the bulk.

At $t=0$, applying the formula (\ref{phiminT2}) we get
\begin{equation}\label{globalphit}
       \phi_{min} + \int_{\Sigma} d\sigma\, \cos \sigma T_{tt}^{M} (\sigma ,t) =    - \sin \sigma \partial_{t}^2 \phi |_{bdy,t=0}  .
\end{equation}
While if we apply the same formula (\ref{phiminT2}) to the case of two decoupled black holes with horizon dilaton field value $\phi_h$ determined by the coupling as in eqn. (\ref{phih}), we get
\begin{equation}\label{tfdphit}
    \phi_h =   - \sin \sigma \partial_{t}^2 \phi |_{bdy,t=0} .
\end{equation}
As we said, after we turn off the coupling, we expect the behavior of $\phi$ near the boundary is close to the profile of dilaton field in the case of two decoupled black holes, with certain horizon value $\phi'_h$. What's more, we know that $\phi'_h$ must be equal to $\phi_h$ in (\ref{tfdphit}), which is required for consistency with the Schwarzian calculation in section 4.3 of \cite{Maldacena:2018lmt}. In section 4.4, we will also check this point by explicit bulk calculation. Then we conclude the boundary terms in (\ref{globalphit}) and (\ref{tfdphit}) are the same, which means the left-hand sides of the two equations are the same:
\begin{equation}
      \phi_{min} + \int_{\Sigma} d\sigma\, \cos \sigma T_{tt}^{M} (\sigma ,t) = \phi_h,
\end{equation}
or
\begin{equation}\label{bhfirst}
    \delta \phi = - \int_{\Sigma} d\sigma\, \cos \sigma T_{tt}^{M} (\sigma ,t) .
\end{equation}

On the other hand, by the entanglement first law, we have
\begin{equation}\label{entfirst}
    \delta S_{bulk} = 2\pi \int_{\Sigma} d\sigma\, \cos \sigma \,T_{tt}^{M} (\sigma ,t) .
\end{equation}

Combining (\ref{bhfirst}) and (\ref{entfirst}), we find
\begin{equation}\label{smarr}
    2\pi\delta \phi +  \delta S_{bulk}=0.
\end{equation}
This is what we found through explicit computation in the last section. However, here the arguments are general and do not require knowing specific matter field content in the bulk. We see that it is a consequence of two first laws: the first law of black hole and the first law of entanglement.

We should note that $\delta S_{bulk} = \delta \langle K\rangle$ only holds approximately. For the case of $\Delta = 1/2$, this holds at linear order of the coupling $\epsilon$ (or $\eta$). For general $\Delta$, since the entanglement entropy is not analytic in $\eta$, $\eta$ is not a good small parameter for doing expansion. In these cases, we should think that $\delta S_{bulk} = \delta \langle K\rangle$ holds at linear order of the temperature $T(\eta)$. In general, we have the inequality $\delta S_{bulk} \leq \delta \langle K\rangle$ \cite{Casini:2008cr}, which means
\begin{equation}
   2\pi \phi_{min} + \delta S_{bulk} \leq  2\pi \phi_h.
\end{equation}
This is saying that, if we fix the boundary behavior of the dilaton field, then the two decoupled black holes with bulk matter fields in the vacuum state has the greatest entanglement between the two sides. Eqn. (\ref{dphids}) simply follows from this maximization condition for small variations.

An analogous statement in field theory is the following. Take two copies of an arbitrary quantum system, with decoupled Hamiltonian $H_L + H_R$. Then among all the pure states $\{\ket{\psi}\}$ with the same total energy as $\ket{TFD(\beta)}$, i.e.
\begin{equation}
    \bra{\psi} H_L + H_R \ket{\psi} = \bra{TFD (\beta)} H_L + H_R \ket{TFD (\beta)},
\end{equation}
the $\ket{TFD(\beta)}$ state itself has maximal entanglement between two sides \footnote{It is not unique though. For example, we can evolve the state using $H_L$, the new state $e^{-i H_L t}\ket{TFD(\beta)}$ has the same entanglement entropy between the two sides.}. This holds because
\begin{equation}
    S_L (\ket{\psi}) - S_L (\ket{TFD(\beta)}) \leq \beta \Delta \langle H_L\rangle,
\end{equation}
and 
\begin{equation}
    S_R (\ket{\psi}) - S_R (\ket{TFD(\beta)}) \leq  \beta\Delta \langle H_R\rangle.
\end{equation}
By adding up the two inequalities, and note that $S_L = S_R$ for pure states, we have
\begin{equation}
    S_L (\ket{\psi}) \leq  S_L (\ket{TFD(\beta)}).
\end{equation}

\subsection{Bulk evolution after turning off the coupling}\label{evolution}

In this section, we study the bulk evolution after we turn off the coupling, and check explicitly that the dilaton field close to the boundary indeed has the form as the TFD solution with the corresponding temperature. We will still study the case of $\Delta =1/2$ as a concrete example, while the qualitative results hold for general $\Delta$. Suppose before $t=0$, the two boundary trajectories are located at $\sigma  = \delta$ and $\sigma = \pi - \delta$, where $\delta \ll 1$. We turn off the coupling suddenly at $t=0$, which creates some positive energy on the boundary. Since the bulk fields are massless, the positive energy will propagate along the null direction into the bulk, forming two shock waves. Thus the stress tensor for $0<t<\pi$ can be written as
\begin{equation}
    T_{++}^{M} (x^+)/N = - \frac{\epsilon}{8\pi} + \frac{\epsilon}{8} \delta (x^+ - \pi + \delta) ,\,\,\,\,   T_{--}^{M} (x^-)/N = - \frac{\epsilon}{8\pi} + \frac{\epsilon}{8} \delta (x^- + \delta) ,
\end{equation}
where we've introduced coordinates $x^\pm = t \pm \sigma$. In these coordinates, the equations of motion for the dilaton field are
\begin{equation}
    \partial_{+} \left(  \sin^2 \sigma \partial_+ \phi   \right) = -T_{++}^{M} \sin^2 \sigma,
\end{equation}
\begin{equation}
    \partial_{-} \left(  \sin^2 \sigma \partial_- \phi   \right) = -T_{--}^{M} \sin^2 \sigma,
\end{equation}
\begin{equation}
     \partial_+ \partial_{-} \phi + \frac{1}{2\sin^2 \sigma} \phi = T_{+-}^{M}.
\end{equation}

\begin{figure*}[t!]
    \centering
    \begin{subfigure}[h!]{0.4\textwidth}
        \centering
        \includegraphics[height=2 in]{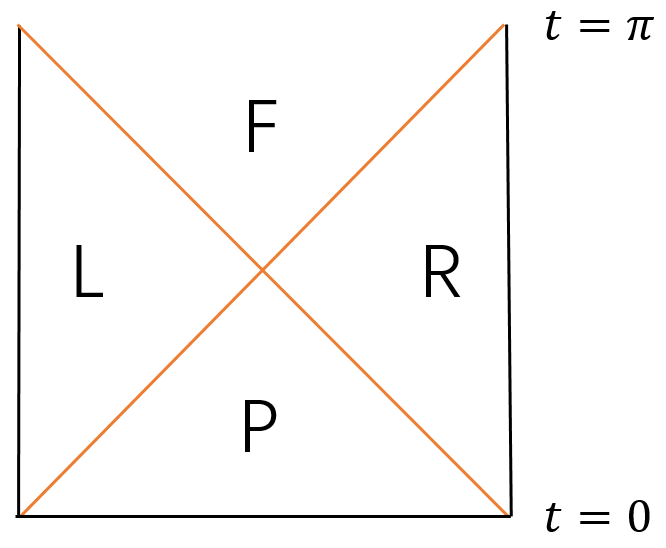}
        \caption{}
        \label{dilatonfig1}
    \end{subfigure}%
    ~ 
    \begin{subfigure}[h!]{0.6\textwidth}
        \centering
        \includegraphics[height=3 in]{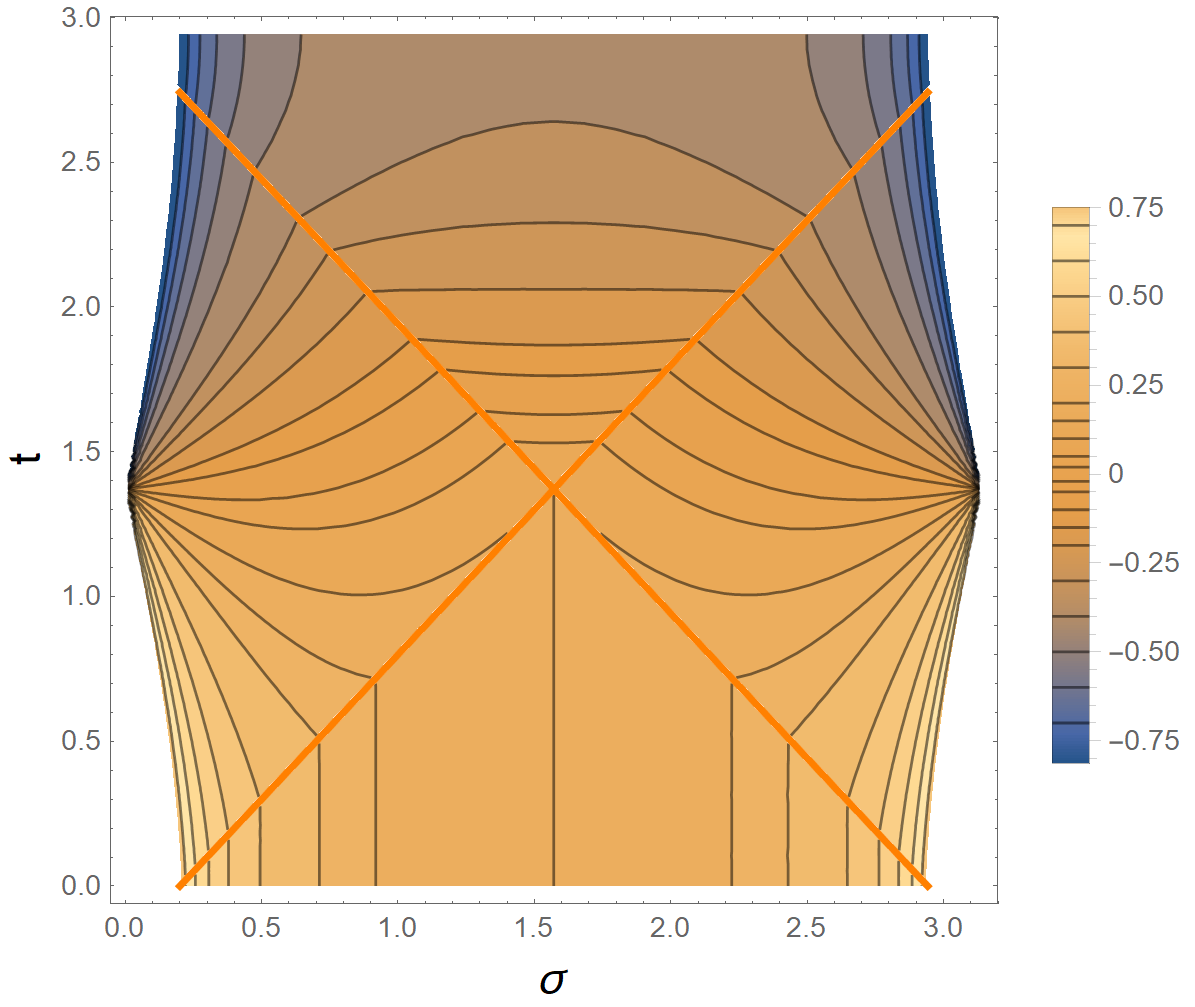}
        \caption{}
        \label{dilatonfig2}
    \end{subfigure}
    \caption{(a) The two shock waves (orange lines) divide the bulk into four regions, which we denote as L (left), R (right), F (future), P (past). (b) The contour plot of the dilaton profile for $t>0$. In the plot, we choose the parameters $\frac{\epsilon}{4\pi} =0.1,\,\delta = 0.2$. The orange lines denote the shock waves.}
\end{figure*}

The shock waves divide the bulk into four regions (see illustration in fig. \ref{dilatonfig1}). We denote the regions as L (left), R (right), F (future), P (past). The derivative of the dilaton field is not continuous when it crosses the shock wave. 

In appendix. \ref{sec:solvedilaton}, we present the calculation that solves the dilaton profile. The results are:
\begin{equation}\label{eqn18}
    \frac{\phi}{N}=\left\{ 
    \begin{aligned}
   & \frac{\epsilon}{4\pi} \left( 1+  \frac{\frac{\pi}{2} - \sigma}{\tan\sigma} \right) , \,\,\,\,\,\,\,\,\,\,\,\,\,\,\,\,\, & (\sigma,t) \in P,\\
   & \frac{\epsilon}{4\pi} \left(1 - \frac{\sigma}{\tan \sigma} + \frac{\pi}{2}\frac{\cos (t +\delta ) }{\sin \sigma}\right), & (\sigma,t) \in L,\\
    & \frac{\epsilon}{4\pi} \left( 1 + \frac{\pi - \sigma}{\tan \sigma} + \frac{\pi}{2}\frac{\cos (t+\delta)}{\sin \sigma}\right), & (\sigma,t) \in R,\\
     & \frac{\epsilon}{4\pi} \left(1+  \frac{\frac{\pi}{2} - \sigma}{\tan\sigma} + \frac{\pi \cos (t+\delta)}{\sin \sigma}\right), & (\sigma,t) \in F,\\
    \end{aligned}
    \right.
\end{equation}

From (\ref{eqn18}) we see that in the left region (L), when we approach the left boundary $\sigma\rightarrow 0$, and for $\delta \ll 1$, we have
\begin{equation}
    \phi_L \approx N\frac{\epsilon}{8}\frac{\cos t}{\sin \sigma},
\end{equation}
which has the form of the solution in the decoupled case, with
\begin{equation}\label{suddenhorizon}
    \phi_h = N\epsilon /8 = N\eta/2.
\end{equation}

The value of $\phi_h$ is consistent with the one in eqn. (\ref{phihdelta1/2}), which was determined by the temperature $T(\eta)$. We also have the same physics near the right boundary. This is consistent with the analysis using the Schwarzian theory. These justify the statements that we made in the last section.

Away from the boundary, the solution has an interesting shape, which is depicted in fig. \ref{dilatonfig2}.

In previous sections including this one, we simplify the problem by considering the case where we suddenly turn off the coupling at $t=-0^+$. More naturally, one would imagine turning off the coupling through an interval of boundary time $u$ of order $\beta$. If we turn off the coupling slower, we should expect to get a state that is close to a TFD state but with slightly lower temperature. In the next section, we study this problem by both gravity calculation and Schwarzian calculation. An interesting point in the calculation is that, to get the correct entanglement entropy on the gravity side, we need to carefully search for the location of the quantum extremal surface.

\subsection{Quench problem of turning off the coupling slowly}\label{sec:slow}

\subsubsection{Gravity calculation}

\begin{figure}[t!]
    \centering
    \includegraphics[width=12cm]{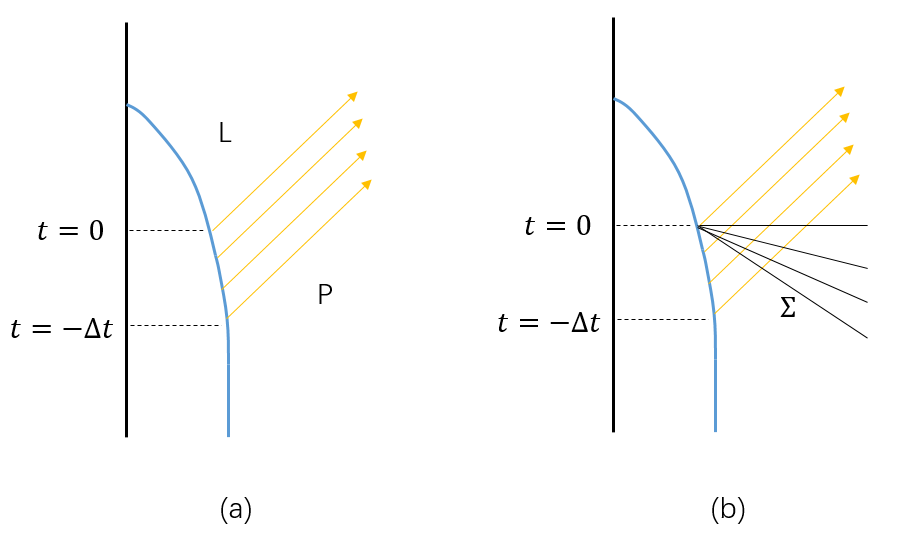}
    \caption{(a) An illustration of the set up. In the figure, only the left part of the spacetime is drawn. We denote the regions before and after the shock waves as P and L. (b) We vary the slice $\Sigma$ with left end fixed at $t=0$, to find the one that maximizes the quantum correction.}
    \label{fig:slow1}
\end{figure}

Imagine we turn off the coupling during time $ t \in [- \Delta t, 0]$. The time scale in terms of the boundary time $u$ is of order $\beta$ set by the coupling. To simplify the calculation, we imagine turning off the coupling uniformly in the global coordinate time $t$. We still consider the case of $\Delta  = 1/2$, where the shock waves travel in the null direction and thus the calculation is simplified. In this case, we can write the energy distribution in the bulk as
\begin{equation}
      T_{++}^{M} (x^+)/N = - \frac{\epsilon}{8\pi} + \frac{\epsilon}{8} f(x^+ -\pi) ,\,\,\,\,   T_{--}^{M} (x^-)/N = - \frac{\epsilon}{8\pi} + \frac{\epsilon}{8} f(x^-) ,
\end{equation}
where $f(x)$ is a ``window" function defined as
\begin{equation}
    f(x) \equiv \left\{ 
    \begin{aligned}
    & \frac{1}{\Delta t},\,\,\,\, & x\in [-\Delta t,0],\\
    & 0, & x \notin [-\Delta t,0].
    \end{aligned}
    \right.
\end{equation}
In fig. \ref{fig:slow1}(a), we show an illustration of the set up. We denote the left region after the shock waves as L. 

By solving the equations of motion of the dilaton as in appendix. \ref{sec:solvedilaton}, we can get the profile of dilaton after the shock waves:
\begin{equation}
  \begin{aligned}
    \phi_L =  N\frac{\epsilon}{4\pi} \left(  1 - \frac{\sigma}{\tan \sigma} + \frac{\pi\sin \frac{\Delta t}{2}}{\Delta t} \frac{\cos (t+ \frac{\Delta t}{2})}{\sin \sigma}  \right).
\end{aligned}
\end{equation}
We can see from the formula that near the boundary $\sigma \sim 0$, $\phi$ behaves like the solution for a black hole, while the horizon of the black hole locates at $t= -\Delta t/2$. The interpretation is that on average we are turning off the coupling at time $t= -\Delta t/2$. The corresponding black hole solution has horizon dilaton field value
\begin{equation}\label{lowerphih}
    \phi_h = N\frac{\epsilon\sin \frac{\Delta t}{2}}{4\Delta t}.
\end{equation}
The value of $\phi_h$ is smaller than the one in (\ref{suddenhorizon}) that we turn off the coupling suddenly. This is simply saying that, by turning off the coupling slower, the state that we get at $t=0$ is close to a thermofield double state with lower temperature. In the next section, by Schwarzian calculation, we can also see this behavior.

We want to calculate the entanglement entropy of the state at $t=0$ using the HRT proposal \cite{Hubeny:2007xt}. By the proposal, the HRT surface should locate at the position that the surface area is extremized. By the maximin proposal of finding HRT surface \cite{Wall:2012uf}, we can first choose a spatial slice $\Sigma$ and find the minimal area surface on that slice, and then vary the slice to find the one with the maximal area. After taking the bulk quantum correction into account, the quantum HRT proposal \cite{Faulkner:2013ana,Engelhardt:2014gca,Dong:2017xht} states we should extremize the sum of the surface area and the quantum correction. The final surface after extremizing is also called the quantum extremal surface. In the following, we'll find the location of the HRT surface through explicit calculation.
By the reflection symmetry between left and right, the HRT surface should locate at $(-t_0 , \pi/2)$, and we vary $t_0$ such that
\begin{equation}\label{extremize}
    \int_{\Sigma } d\sigma \sqrt{h} T_{\mu\nu}^{M} n^\mu \xi^\nu + \phi \left(-t_0 , \frac{\pi}{2}\right). 
\end{equation}
is maximized. In the limit of $\Delta t \rightarrow 0$ as discussed in previous sections, we have $t_0 =0$. As we can see from the following calculation, when we increase $t_0$ from zero, the sum in (\ref{extremize}) decreases, thus we expect the maximum locates in region P of fig. \ref{fig:slow1}(a). Since the dilaton value is constant along $\sigma = \pi/2$ in this region, we just need to maximize the quantum correction part. In formula (\ref{extremize}), the Killing vector $\xi^{\mu}$ should be the one with fixed point at $(-t_0 , \pi/2)$. This idea of maximization is illustrated in fig. \ref{fig:slow1}(b).

We can calculate the quantum correction explicitly. On the left half of the bulk, we have
\begin{equation}
    T_{tt}^{M} = T_{\sigma\sigma}^{M} = -N\frac{\epsilon}{4\pi} + N\frac{\epsilon}{8}f(t-\sigma),\quad T_{t\sigma}^{M} = T_{\sigma t}^{M} = - N\frac{\epsilon}{8}f(t-\sigma).
\end{equation}

The Killing vector with the corresponding fixed point is
\begin{equation}
    \xi^t = \cos \sigma \cos (t + t_0),\,\,\,\, \xi^\sigma = - \sin \sigma \sin (t+ t_0).
\end{equation}

We choose coordinate $\sigma$ to parametrize the slice $\Sigma$, and on the slice we have $t = - (2t_0/\pi)\sigma$. The induced metric and the normal vector of slice $\Sigma$ are
\begin{equation}
    \sqrt{h} = \sqrt{\frac{1- \left( \frac{2t_0}{\pi}  \right)^2}{\sin^2 \sigma}},\quad n^{\mu} =\frac{1}{ \sqrt{\frac{1- \left( \frac{2t_0}{\pi}  \right)^2}{\sin^2 \sigma}}} \left\{   1  , - \frac{2t_0}{\pi}  \right\}.
\end{equation}

Then we have
\begin{equation}
    \begin{aligned}
   &  \quad \quad \int_{\Sigma } d\sigma \sqrt{h} T_{\mu\nu}^{M} n^\mu \xi^\nu \\
   & = N\int_{0}^{\frac{\pi}{2}}d\sigma \, \left( -\frac{\epsilon}{4\pi} + \frac{\epsilon}{8}f\left(- \frac{2t_0}{\pi}\sigma-\sigma\right) \right)\cos \sigma \cos \left(- \frac{2t_0}{\pi}\sigma + t_0\right)  \\
     & \quad  + N\int_{0}^{\frac{\pi}{2}}d\sigma \, \left( -\frac{\epsilon}{4\pi} + \frac{\epsilon}{8}f\left(- \frac{2t_0}{\pi}\sigma-\sigma\right) \right)\left(-\sin \sigma \sin \left(- \frac{2t_0}{\pi}\sigma + t_0\right)\right) \left(-\frac{2t_0}{\pi}\right) \\
      &  \quad + N\int_{0}^{\frac{\pi}{2}}d\sigma \, \left( - \frac{\epsilon}{8}f\left(- \frac{2t_0}{\pi}\sigma-\sigma\right) \right)\left(-\sin \sigma \sin \left(- \frac{2t_0}{\pi}\sigma + t_0\right)\right) \\
      & \quad  + N\int_{0}^{\frac{\pi}{2}}d\sigma \, \left( - \frac{\epsilon}{8}f\left(- \frac{2t_0}{\pi}\sigma-\sigma\right) \right)\cos \sigma \cos \left(- \frac{2t_0}{\pi}\sigma + t_0\right) \left(-\frac{2t_0}{\pi}\right) \\
      & =  -N\frac{\epsilon}{4\pi}  + N\frac{\epsilon}{8} \int_{0}^{\frac{\Delta t}{1+ 2t_0 /\pi}} d\sigma \,  \left( 1+ \frac{2t_0}{\pi}  \right)  \cos\left( \sigma + \frac{2t_0}{\pi} \sigma - t_0 \right) \\
      & =  -N\frac{\epsilon}{4\pi} + N\frac{\epsilon}{8} \left(  \sin (\Delta t - t_0 ) + \sin t_0  \right).
    \end{aligned}
\end{equation}
By maximizing 
\begin{equation}
     \sin (\Delta t - t_0 ) + \sin t_0 ,
\end{equation}
we find that we should choose $t_0 = \Delta t/2$. This means that the location of the HRT surface is the same as the location of the horizon for the corresponding black hole solution.

\begin{figure}[t!]
    \centering
    \includegraphics[width=12cm]{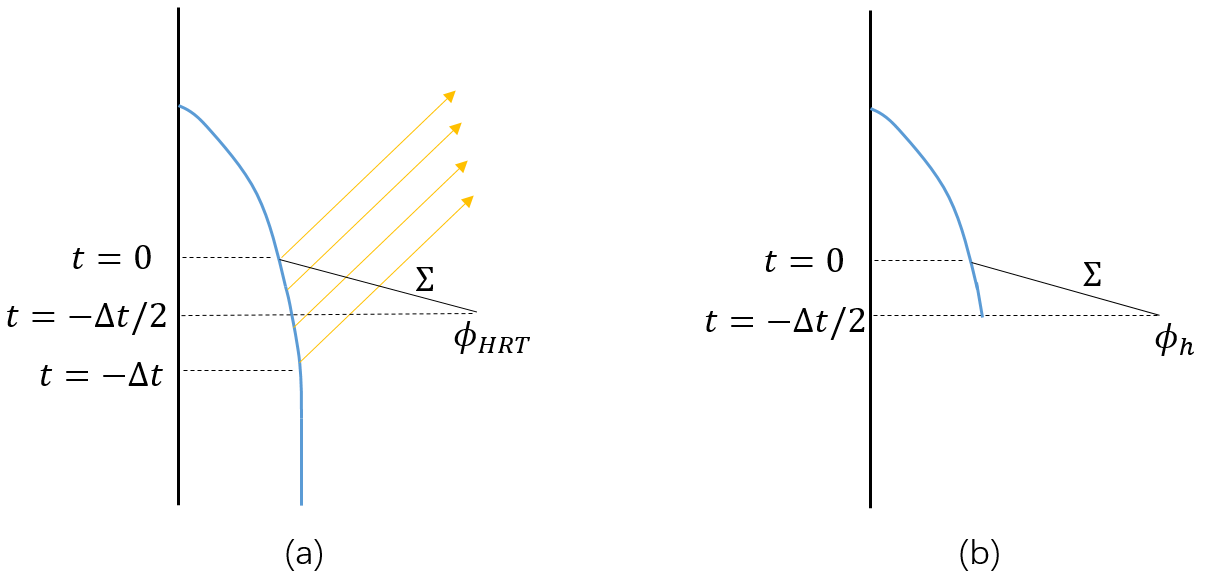}
    \caption{(a) The scenario that we turn off the coupling slowly. The horizon locates at $t= - \Delta t/2$. (b) The corresponding black hole solution with horizon located at $t=-\Delta t/2$.}
    \label{fig:slow3}
\end{figure}

We denote the dilaton field value at the HRT surface as $\phi_{HRT}$. As we've derived in sec. \ref{sec:fisrtlawformula}, we can relate $\phi_{HRT}$ and the quantum correction to a boundary term. The specific form of the boundary term is not important here. When we apply the formula (\ref{covariant}) to fig. \ref{fig:slow3}(a), we get
\begin{equation}
    \phi_{HRT} +   \int_{\Sigma } d\sigma \sqrt{h} T_{\mu\nu}^M n^\mu \xi^\nu = \textrm{boundary term at $t=0$} .
\end{equation}
While if we apply the formula (\ref{covariant}) to fig. \ref{fig:slow3}(b), since the quantum correction part is zero, we get
\begin{equation}
     \phi_{h}= \textrm{boundary term at $t=0$} .
\end{equation}
The two boundary terms are the same since $\phi$ is the same in the two figures close the boundary for $t>0$. So we get
\begin{equation}\label{13}
     \phi_{HRT} +   \int_{\Sigma } d\sigma \sqrt{h} T_{\mu\nu}^M n^\mu \xi^\nu  = \phi_h.
\end{equation}
This is saying that if we turn off the coupling slowly, the entanglement entropy of the state that we get at $t=0$ is the same as that of the thermofield double state with lower temperature determined by (\ref{lowerphih}).

\subsubsection{Schwarzian analysis}

In last section, we analyzed the time evolution after we slowly turn off the interaction, from the bulk point of view. In this section, we discuss the same physics using the boundary Schwarzian theory. We start from action:
\begin{align}
S=N\int du\, \left[ -\left\{\tan \frac{t_l(u)}{2},u\right\}-\left\{\tan \frac{t_r(u)}{2},u\right\}+\eta(u)\left(\frac{t_l'(u)t_r'(u)}{\cos^2(\frac{t_l-t_r}{2})}\right)^\Delta \right],
\end{align}
where $\eta(u)$ is the time-dependent left-right coupling strength. In the bulk calculation in previous section, we assumed that the coupling is turned off with constant speed, with respect to the coordinate time $t$. Here instead we assume that the coupling is uniformly turned off with repect to the boundary time $u$:
\begin{align}
\eta(u<-\delta u)=\eta,\ \ \ \ \ \ \ \ \eta(-\delta u <u<0)=-\frac{\eta u}{\delta u},\ \ \ \ \ \ \ \ \eta(u>0)=0.
\end{align}
We will look at the region where $\delta u$ is much smaller than the time scale set by the coupling and work perturbatively. The results should match the bulk calculation in the previous section when we look at the leading order results.

As in \cite{Maldacena:2018lmt}, it's convenient to make a transformation of variables: 
\begin{align}
t_l(u)=t_r(u)=t(u), \ \ \ \ t'(u)=\exp(\varphi(u)).
\end{align}
Then the equation of motion of the Schwarzian theory can be written as:
\begin{align}\label{schwarzianphi}
\varphi''(u)+e^{2\varphi(u)}-\eta(u)\Delta e^{2\Delta \varphi(u)}=0.
\end{align}

For $-\delta u <u<0$, the evolution is determined by:
\begin{align}
\varphi''(u)+e^{2\varphi(u)}+\Delta \frac{\eta u}{\delta u}e^{2\Delta \varphi(u)}=0, 
\end{align}
with boundary conditions:
\begin{equation}
    \varphi(-\delta u)=\frac{1}{2(1-\Delta)}\log(\eta \Delta), \quad  \varphi'(-\delta u)=0.
\end{equation}
For $u>0$, since we have $\eta(u) =0$, the solution of (\ref{schwarzianphi}) is the solution corresponding to a TFD state. There are two parameters of the general solution, one is the temperature, and the other is the time shift. To determine the solution, we only need to find the boundary conditions $\varphi(0)$ and $\varphi'(0)$. They can be expanded in terms of $\delta u$. To the leading order, the equation for $\delta \varphi(u)=\varphi(u)-\varphi(-\delta u)$ is given by:
\begin{align}
\delta \varphi''+(\Delta \eta)^\frac{\Delta}{1-\Delta}\frac{u \eta}{\delta u}+(\Delta \eta)^\frac{1}{1-\Delta}=0,
\end{align}
which leads to the solution:
\begin{align}
\delta \varphi(u)=-\frac{1}{6\delta u}(\delta u+u)^3(\Delta \eta)^{\frac{1}{1-\Delta}}.
\end{align}
Then the boundary condition at $u=0$ is:
\begin{align}
\delta \varphi'(0)=-\frac{1}{2}\delta u(\Delta \eta)^\frac{1}{1-\Delta},\ \ \ \ \delta \varphi(0)=-\frac{1}{6}\delta u^2(\Delta \eta)^\frac{1}{1-\Delta}.
\end{align}
The general form of the solution for $u>0$ is
\begin{align}
\exp(\varphi(u))=\frac{a}{\cosh(a u+b)},\ \ \ \ t(u)=t_0+2\arctan\left[\tanh\frac{a u+b}{2}\right].
\end{align}
Parameter $a$ is related to the temperature $T$ of the TFD state by $a=2\pi T$. Parameter $b$ measures the time shift. For a sudden quench with $\delta u=0$, we have $b_0=0$ and $a_0=(\eta\Delta)^{\frac{1}{2(1-\Delta)}}$. For finite but small $\delta u$, the leading order results are
\begin{equation}\label{dectemp}
    a=a_0 \left(1-\frac{\delta u^2(\eta \Delta)^{\frac{1}{1-\Delta}}}{24}\right),
\end{equation}
\begin{equation}\label{shifttime}
    b=a_0\frac{\delta u}{2}.
\end{equation}
The first formula (\ref{dectemp}) tells us that if we turn off the coupling slower, we get a TFD state with lower temperature. For the special case of $\Delta=1/2$, we have $T=T_0(1-\delta u^2\eta^2 /96)$, where $T_0$ is the temperature corresponding to a sudden quench. We can compare this result with the bulk calculation. In the bulk calculation where we turn off the coupling during time $\delta t$, we found that the dilaton field at the horizon has value
\begin{equation}
    \phi \propto \frac{\sin(\delta t/2)}{\delta t/2}\sim1-\frac{\delta t^2}{24} = 1-\frac{\delta u^2\eta^2}{96},
\end{equation}
where we used $t=t' u$ and $t'= \eta/2$. We found the two results match with each other as expected.

The formula (\ref{shifttime}) is saying that at $u=0$, we in fact get a TFD state with time shift $\delta u/2$. The same is found in the bulk calculation, where we've seen that the horizon locates at $t= - \delta t/2$.

\section{Conclusion and Discussion}

In this paper, we studied the entanglement entropy in the ground state of the coupled SYK system. We found that for small coupling, the entanglement entropy in the ground state $\ket{G(\eta)}$ is the same as the entanglement entropy in the thermofield double state $\ket{TFD(\eta)}$. This is consistent with the claim that the two states are close to each other. 

We illustrated how to calculate the entanglement entropy by using the Ryu-Takayanagi formula in gravity. Each bulk field gives small correction to the entanglement entropy, however, the effect is enhanced to be of order $N$ due to the large number of bulk fields. At first sight, the gravity pictures for $\ket{G(\eta)}$ and $\ket{TFD(\eta)}$ look very different. In one case, we have an eternal traversable wormhole, and in the other case, we have two decoupled black holes. However, after we add up the RT surface term and the bulk quantum correction term, we find the two states have the same entanglement entropy. From the gravity point of view, this is guaranteed by two different first laws: the first law of black holes, and the first law of entanglement. The linearity of dynamics in JT gravity allows us to apply the first law to two seemingly very different solutions. In some sense, the two-sided black hole situation (or the TFD state) is special as it has maximal entanglement between two sides as we fix the total energy.  

We also discussed the gravitational dynamics of turning off the coupling suddenly or slowly. For the case of turning off the coupling slowly, to apply the HRT formula, we need to search for the quantum extremal surface. The simplicity of near-AdS$_2$ gravity allows us to do the calculation explicitly. We found agreements between the bulk calculation and the boundary Schwarzian calculation. 

There are some interesting questions to explore in the future. One can apply the formula (\ref{covariant}) to other Killing vectors that have fixed points on places that the dilaton is not extremal. Then it will relate the dilaton field on the new fixed point and the Rindler energy associated with the new Killing vector, to a different boundary term. We expect the boundary term will correspond to the SL(2,R) charge of the boundary trajectory associated with the Killing vector. However, it is unclear how to interpret these family of formulas since in these cases the dilaton field is not the RT surface. It would be nice to understand their physical interpretation in the SYK model.

In \cite{Callebaut:2018nlq,Callebaut:2018xfu}, instead of viewing the dilaton field as the RT surface for calculating entanglement entropy in the SYK model, the authors view the JT gravity as emergent from the entanglement dynamics in a 2D CFT with boundary. It would be nice to understand the connection between these two viewpoints better.

In this paper, we only studied the entanglement entropy in the ground state of the coupled system. It is also interesting to study the entanglement entropy (or R\'{e}nyi entropy) at finite temperature. As discussed in \cite{Maldacena:2018lmt}, at finite temperature, the coupled system has a Hawking-Page like phase transition in the canonical ensemble, while there is no phase transition in the microcanonical ensemble. This implies that in the canonical ensemble, the gravity geometry has a sudden topology change, while the system has to go through a non-geometric phase in the microcanonical ensemble. It would be interesting to see how the entanglement behaves in these situations.

\paragraph{Acknowledgement}

\begin{appendices}
We want to thank Juan Maldacena for patient guidance and encouragement through the project. We also want to thank Xi Dong, Tarun Grover, Yingfei Gu, Ho Tat Lam, Henry Lin, Xiao-Liang Qi, Cenke Xu, Zhenbin Yang and Hui Zhai for helpful discussions. PZ was supported in part by the National Science Foundation under Grant No. NSF PHY-1748958 and the Heising-Simons Foundation.

\section{Energy of CFT on a strip with varying boundary couplings}\label{sec:energyCFT}

We consider a conformal field theory on an infinite strip $[0,\pi] \times \mathbb{R}$. We add a deformation $\int g(t)O(t,0)O(t,\pi)$ to the CFT, which couples the two boundaries directly through primary operator $O$. We would like to calculate the expectation value of the stress tensor.

We begin with the conformal OPE
\begin{equation}
    \langle  T_{zz}(z) O(t_1,0)O(t_1,\pi)  \rangle = -\frac{\Delta}{(2\pi) 2^{2\Delta}} \frac{1}{\sinh^2 (t-t_1 + i\sigma)},
\end{equation}
 By analytic continuation to Lorentzian signature, we get
\begin{equation}\label{T++}
    \langle T_{++}(x^+) O(t_1, 0)O(t_1,\pi) \rangle =  \frac{\Delta}{(2\pi) 2^{2\Delta}} \frac{1}{\sin^2 (x^+ - t_1)},
\end{equation}
where we've introduced light cone coordinates $x^\pm = t \pm \sigma$. The expectation value is taken in the vacuum state on the strip.

The expression in eqn. (\ref{T++}) has poles at $x^+ \sim t_1 + m \pi$. Expanding it around a pole, we get
\begin{equation}
    \langle   T_{++}(x^+) O(t_1, 0)O(t_1,\pi)  \rangle  \sim  -\frac{\Delta}{(2\pi) 2^{2\Delta}} \partial_{x^+} \frac{1}{x^+ -t_1 - m\pi},
\end{equation}
and thus we have
\begin{equation}
     \langle   [T_{++}(x^+), O(t, 0)O(t,\pi)]  \rangle  =  -i\frac{\Delta}{ 2^{2\Delta}}\sum_{m= -\infty}^{\infty}\delta' (x^+ - t - m\pi).
\end{equation}
Similarly, we have
\begin{equation}
       \langle   [T_{--}(x^-), O(t, 0)O(t,\pi)]  \rangle   = -i\frac{\Delta}{ 2^{2\Delta}}\sum_{m= -\infty}^{\infty}\delta' (x^- - t - m\pi).
\end{equation}

We imagine turning on the coupling adiabatically. Then by perturbation theory, we have
\begin{equation}
\begin{aligned}
    \langle T_{++} (x^+) \rangle_g|_{t=0} & = i\int_{-\infty}^{0} g(t) \langle   [T_{++}(x^+), O(t, 0)O(t,\pi)]  \rangle  + \mathcal{O}(g^2) \\
    & =\frac{\Delta}{ 2^{2\Delta}} \sum_{m= -\infty}^{\infty}\int_{-\infty}^{0} g(t) \delta' (x^+ - t - m\pi)  , 
\end{aligned}
\end{equation}
\begin{equation}
\begin{aligned}
    \langle T_{++} (\sigma,t=0) \rangle_g  & = \frac{\Delta}{ 2^{2\Delta}}\sum_{m= -\infty}^{\infty}\int_{-\infty}^{0} g(t) \delta' (\sigma - t - m\pi) =-\frac{\Delta}{ 2^{2\Delta}} \sum_{m=1}^{\infty} g'(\sigma - m\pi),
\end{aligned}
\end{equation}
and
\begin{equation}
\begin{aligned}
    \langle T_{--} (\sigma,t=0) \rangle_g  & = \frac{\Delta}{ 2^{2\Delta}} \sum_{m= -\infty}^{\infty}\int_{-\infty}^{0} g(t) \delta' (-\sigma - t - m\pi) =- \frac{\Delta}{ 2^{2\Delta}} \sum_{m=0}^{\infty} g'(-\sigma - m\pi).
\end{aligned}
\end{equation}
Note by $\langle\rangle_g$, we mean the expectation value with finite $g(t)$ substract the vacuum expectation value ($g(t)=0$), thus the vacuum Casimir energy piece cancels. 

For the process described in sec. \ref{sec:quantumcorrection}, we start from $g(t)=0$ at $t=-\infty$, and we adiabatically turn on $g(t)$ to a constant value $g$, then at $t=-0^+$ right before $t=0$, we suddenly turn off the coupling. We can calculate the energy distribution in the bulk as follows:

\begin{equation}
\begin{aligned}
    \langle  T_{++} (\sigma,t=0)  \rangle_g = - \frac{\Delta}{ 2^{2\Delta}}\sum_{m=1}^{\infty} g'(\sigma - m\pi)  =\frac{\Delta g}{ 2^{2\Delta}}\left[ -\frac{1}{\pi} +  \, \delta (\sigma- \pi + 0^+)\right],
\end{aligned}
\end{equation}
\begin{equation}
\begin{aligned}
    \langle  T_{--} (\sigma,t=0)  \rangle_g = - \frac{\Delta}{ 2^{2\Delta}}\sum_{m=0}^{\infty} g'(\sigma - m\pi)  = \frac{\Delta g}{ 2^{2\Delta}}\left[-\frac{1}{\pi} + \, \delta (\sigma - 0^+)\right],
\end{aligned}
\end{equation}
where we have assumed that the increase of $g(t)$ is slow enough to approximate the summation by integration.

For the massless fermion case with $\Delta = 1/2$ that was discussed in sec. \ref{sec:quantumcorrection}, we have
\begin{equation}\label{Ttt}
    \langle  T_{tt} (\sigma,t=0)  \rangle_{\epsilon} =  \langle  T_{++} (\sigma,t=0)  \rangle_{\epsilon} +  \langle  T_{--} (\sigma,t=0)  \rangle_{\epsilon}  = -\frac{\epsilon}{4\pi} + \frac{\epsilon}{8} \left(\delta (\sigma- 0^+)  + \delta (\sigma-\pi + 0^+)\right),
\end{equation}
where we have fixed the proportional constant by comparing to eqn. (\ref{energydis}).

\section{Direct calculation of the quantum correction}\label{sec:directcorrection}

In this section, we use correlation functions to calculate the entanglement entropy and the modular Hamiltonian for a free fermion system on a strip. The method is introduced in \cite{Peschel_2003}. We consider a real massless fermion living on the strip $[0,\pi] \times \mathbb{R}$, which is composed of two components $\psi_{+},\psi_{-}$. The case discussed in sec. \ref{sec:rteternal} is simply $N$ copies of such system. When we add boundary couplings, we change the boundary conditions of the fermion fields as in eqn. (\ref{bdycondition}).

We start with the mode expansion of the fermion operators:
\begin{equation}
\begin{aligned}
    \psi_+ &  = \sum_{\omega}\left( A_{\omega,+} e^{i\omega (x-t)} b_\omega +   A_{\omega,+}^{*} e^{-i\omega(x-t)} b_\omega^{\dagger} \right),\\
    \psi_- & = \sum_\omega \left( A_{\omega,-} e^{-i\omega (x+t)} b_\omega  + A_{\omega,-}^{*} e^{i\omega (x+t)}b_\omega^{\dagger}   \right).
\end{aligned}
\end{equation}
From the boundary condition in (\ref{bdycondition}), one gets $\omega = \frac{1}{2}+ \epsilon + 2n$ for $n\geq 0 $ and $\omega =-\frac{1}{2}-\epsilon + 2n $ for $n\geq 1$, and
\begin{equation}
    A_{\omega_-} = \frac{1+\sin\pi \epsilon e^{i\omega\pi}}{ \cos \pi \epsilon } A_{\omega, +} \equiv \gamma_{\omega} A_{\omega,+}.
\end{equation}
We normalize $A_+$ and $A_-$ as $|A_+| = |A_-| = \frac{1}{\sqrt{2\pi}}$.
We have $\gamma_\omega = e^{i\epsilon \pi}$ for $\omega = \frac{1}{2}+ \epsilon + 2n$ and $\gamma_\omega = e^{-i\epsilon \pi}$ for $\omega = -\frac{1}{2}- \epsilon + 2n$.

The equal time correlation functions in the ground state are calculated as below:
\begin{equation}
    \langle  \psi_{+}(x) \psi_{+}(y)  \rangle = \sum_{\omega} A_{\omega,+}A_{\omega,+}^{*} e^{i\omega (x-y+i0+)} \langle b_\omega b^{\dagger}_{\omega}  \rangle = \sum_{\omega} \frac{1}{2\pi}  e^{i\omega (x-y+i0+)} =  \frac{i \cos \left[ (\frac{1}{2} - \epsilon) (x-y) \right]}{2\pi \sin (x-y + i0+)},
\end{equation}
\begin{equation}
    \langle  \psi_{-}(x) \psi_{-}(y)  \rangle = \sum_{\omega} A_{\omega,-}A_{\omega,-}^{*} e^{i\omega (y-x+i0+)} \langle b_\omega b^{\dagger}_{\omega}  \rangle = \sum_{\omega} \frac{1}{2\pi} e^{i\omega (y-x+i0+)}=\frac{i \cos \left[ (\frac{1}{2} - \epsilon) (y-x) \right]}{2\pi \sin (y-x + i0+)},
\end{equation}
\begin{equation}
    \langle  \psi_{+}(x) \psi_{-}(y)  \rangle = \sum_{\omega} A_{\omega,+}A_{\omega,-}^{*} e^{i\omega (y+x)} \langle b_\omega b^{\dagger}_{\omega}  \rangle = \sum_{\omega} \frac{1}{2\pi}\gamma_\omega^{*} e^{i\omega (x+y)}=\frac{-i  \sin \left[ (\frac{1}{2} - \epsilon) (x+y-\pi) \right]}{2\pi \sin (x+y)},
\end{equation}
\begin{equation}
    \langle  \psi_{-}(x) \psi_{+}(y)  \rangle = \sum_{\omega} A_{\omega,-}A_{\omega,+}^{*} e^{-i\omega (y+x)} \langle b_\omega b^{\dagger}_{\omega}  \rangle = \sum_{\omega} \frac{1}{2\pi}\gamma_\omega e^{-i\omega (x+y)}=\frac{i  \sin \left[ (\frac{1}{2} - \epsilon) (x+y-\pi) \right]}{2\pi \sin (x+y)}.
\end{equation}

We divide the strip $[0,\pi]$ into two halves $[0,\frac{\pi}{2})\cup(\frac{\pi}{2},\pi]$. For convenience, we make a rotation in the space of the fermion operators:  $\psi_1 \equiv \frac{1}{\sqrt{2}}(\psi_+ + \psi_-),\, \psi_2 \equiv \frac{1}{\sqrt{2}}(\psi_+ - \psi_-) $, then the correlation matrix for the left half of the system has the following form:

\begin{equation}
    C(x,y) = \begin{pmatrix}
    \frac{1}{2}\delta (x-y) & \frac{i}{2\pi}A(x,y)\\
    \frac{i}{2\pi}B(x,y) & \frac{1}{2}\delta (x-y)
    \end{pmatrix}, \,\,\,\, x,y\in \left(0,\frac{\pi}{2}\right),
\end{equation}
where we have defined
\begin{equation}
    A(x,y) \equiv \frac{\cos \left[ (\frac{1}{2} - \epsilon) (x-y) \right]}{\sin (x-y)} + \frac{  \sin \left[ (\frac{1}{2} - \epsilon) (x+y-\pi) \right]}{ \sin (x+y)},
\end{equation}
\begin{equation}
    B(x,y) \equiv \frac{\cos \left[ (\frac{1}{2} - \epsilon) (x-y) \right]}{\sin (x-y)} - \frac{  \sin \left[ (\frac{1}{2} - \epsilon) (x+y-\pi) \right]}{ \sin (x+y)}.
\end{equation}

The entanglement entropy between left and right is related to the correlation matrix by \cite{Peschel_2003,Casini:2009vk}: 
\begin{equation}\label{peschelformula}
    S_{EE}= - \frac{1}{2} \textrm{tr}\left[ C\log C + (1-C)\log (1-C) \right],
\end{equation}
where the factor $1/2$ comes from the fact that we are dealing with real fermions instead of complex fermions.

Since we've already calculated the correlation functions for arbitrary $\epsilon$, by taking into (\ref{peschelformula}) we can find out the entanglement entropy exactly for any $\epsilon$. While for our purpose in the main part of this paper, we want to find out the correction of entanglement entropy coming from small non-zero $\epsilon$ and we'll do this by perturbation theory. As a first step, we need to find the eigenvalues and eigenfunctions of $C_0 (x,y)$ (where subscript $0$ means $\epsilon=0$).

When $\epsilon = 0$, we have \begin{equation}
\begin{aligned}
    A_0 (x,y)  = \frac{1}{2}\left( \frac{1}{\sin \frac{x-y}{2} } -  \frac{1}{\sin \frac{x+y}{2}} \right), \,\,\,\,
    B_0 (x,y)  = \frac{1}{2}\left( \frac{1}{\sin \frac{x-y}{2} } + \frac{1}{\sin \frac{x+y}{2}} \right).
\end{aligned}
\end{equation}
For them, we have the following identities:
\begin{equation}
    \dashint_{0}^{\frac{\pi}{2}} dy \, A_0 (x,y)\, \Im \left\{\frac{1}{\left[\sin \frac{y+\frac{\pi}{2}}{2}  \right]^{\frac{1}{2}+is} \left[\sin \frac{\frac{\pi}{2}-y}{2}  \right]^{\frac{1}{2}-is} }   \right\} = \Re \left\{\frac{\pi \tanh \pi s}{\left[\sin \frac{x+\frac{\pi}{2}}{2}  \right]^{\frac{1}{2}+is} \left[\sin \frac{\frac{\pi}{2}-x}{2}  \right]^{\frac{1}{2}-is} }   \right\},
\end{equation}
\begin{equation}
    \dashint_{0}^{\frac{\pi}{2}} dy \, B_0 (x,y)\, \Re \left\{\frac{1}{\left[\sin \frac{y+\frac{\pi}{2}}{2}  \right]^{\frac{1}{2}+is} \left[\sin \frac{\frac{\pi}{2}-y}{2}  \right]^{\frac{1}{2}-is} }   \right\} = \Im \left\{\frac{-\pi \tanh \pi s}{\left[\sin \frac{x+\frac{\pi}{2}}{2}  \right]^{\frac{1}{2}+is} \left[\sin \frac{\frac{\pi}{2}-x}{2}  \right]^{\frac{1}{2}-is} }   \right\},
\end{equation}
where $\dashint$ means Cauchy principal value integral, $\Re, \Im$ denote the real part and the imaginary part. One can check these identities using numerical integration. Define the real part as $f_s$, the imaginary part as $g_s$, we have
\begin{equation}
  \int_{0}^{\frac{\pi}{2}} dy\,  C_0(x,y) \begin{pmatrix} 
    f_s(y) \\ i g_s(y)\end{pmatrix} = \left(\frac{1}{2} - \frac{1}{2}\tanh \pi s  \right)\begin{pmatrix} 
    f_s(x) \\ i g_s(x)\end{pmatrix},\,\,\,\, s\in \mathbb{R}.
\end{equation}
The eigenfunctions can be normalized by
\begin{equation}
    v_s(x) \equiv \frac{1}{\sqrt{2\pi}}\begin{pmatrix} 
    f_s(x) \\ i g_s(x)\end{pmatrix},\,\,\,\,\textrm{s.t.}\,\,\,\, \int_{0}^{\frac{\pi}{2}} dx\, v_s(x)^{\dagger} v_{s'} (x) = \delta (s-s').
\end{equation}

Denote the eigenvalues as $\lambda_s$, one has the following expression for the correction of entanglement entropy:
\begin{equation}\label{entropyshift}
    \delta_{\epsilon} S_{EE} =  \int_{-\infty}^{\infty} ds\, \pi s \,\delta_{\epsilon}\lambda_s ,
\end{equation}
where $\delta_\epsilon \lambda_s$ is given by
\begin{equation}\label{eigenshift}
\begin{aligned}
    \delta_\epsilon \lambda_s &  = \int\int dx  dy \, v_s (x)^{\dagger} \delta_{\epsilon}C(x,y) v_s (y)\\
    & =- \frac{\epsilon}{4\pi^2} \int\int dx  dy \, f_s (x) \left(  \frac{x-y}{\cos\frac{x-y}{2}  } + \frac{x+y-\pi}{\sin \frac{x+y-\pi}{2}}  \right) g_s (y).
\end{aligned}
\end{equation}
Taking (\ref{eigenshift}) into (\ref{entropyshift}), by first doing the integral for $s$, and then doing the integration on $x$ and $y$, we find at leading order of $\epsilon$
\begin{equation}
    \delta_\epsilon S_{EE} = \left(  \frac{\pi}{4} - \frac{1}{2}  \right)\epsilon.
\end{equation}

In the discussion of section \ref{sec:quantumcorrection}, since there are $N$ copies of fermions in the bulk, the total bulk quantum correction of entanglement entropy is $N \left(  \frac{\pi}{4} - \frac{1}{2}  \right)\epsilon$. This agrees with the result in eqn. (\ref{qc}) by using entanglement first law.

With the eigenfunctions and eigenvalues of $C_0(x,y)$, we can also calculate the modular Hamiltonian $\mathcal{H}$ for $\epsilon =0$ explicitly. The calculation is similar to the one in \cite{Arias:2018tmw}. The modular Hamiltonian can be written as 
\begin{equation}\label{mh1}
    \mathcal{H} = \frac{1}{2}\int_{0}^{\frac{\pi}{2}}\int_{0}^{\frac{\pi}{2}} dxdy\, \psi_i(x) H_{ij}(x,y) \psi_j(y),\,\,\,\, i,j = 1,2.
\end{equation}
The matrix $H(x,y)$ can be expressed in terms of the eigenfunctions:
\begin{equation}
    H(x,y) = \int_{-\infty}^{\infty} ds\,  v_{s}(x)2\pi s\, v_{s}(y)^{\dagger}.
\end{equation}
The integration can be carried out:
\begin{equation}
 H_{11}(x,y) =    \int_{-\infty}^{\infty} \, ds \, s f_s(x) f_{s} (y) = 0,\,\,\,\,  H_{22}(x,y) =\int_{-\infty}^{\infty} \, ds \, s \, g_s(x) g_{s} (y) =0,
\end{equation}

\begin{equation}\label{mh0}
\begin{aligned}
 H_{12}(x,y) & = -i  \int_{-\infty}^{\infty} \, ds \, s f_s(x) g_{s} (y) \\
 & =-i \int_{-\infty}^{\infty} \, ds \, s\, \Re \left\{\frac{1}{\left[\sin \frac{x+\frac{\pi}{2}}{2}  \right]^{\frac{1}{2}+is} \left[\sin \frac{\frac{\pi}{2}-x}{2}  \right]^{\frac{1}{2}-is} }   \right\} \Im \left\{\frac{1}{\left[\sin \frac{y+\frac{\pi}{2}}{2}  \right]^{\frac{1}{2}+is} \left[\sin \frac{\frac{\pi}{2}-y}{2}  \right]^{\frac{1}{2}-is} }   \right\} \\
 & = -i\int_{-\infty}^{\infty} \, ds \, k(x,y) \frac{d}{d\omega (x)} \Im \exp\left(is\omega(x)\right)  \Im  \exp\left(is\omega(y)\right)\\
 & =-i\pi k(x,y)\frac{d}{d\omega(x)}\delta(\omega(x)-\omega(y)) = -i\pi k(x,y)\left[ \frac{1}{\omega'(x)^2}\delta'(x-y)  - \frac{\omega''(x)}{\omega'(x)^3}\delta(x-y)  \right],
\end{aligned}
\end{equation}
where we've defined
\begin{equation}
    k(x,y) \equiv  \frac{1}{\left[\sin \frac{x+\frac{\pi}{2}}{2}  \right]^{\frac{1}{2}} \left[\sin \frac{\frac{\pi}{2}-x}{2}  \right]^{\frac{1}{2}} }\frac{1}{\left[\sin \frac{y+\frac{\pi}{2}}{2}  \right]^{\frac{1}{2}} \left[\sin \frac{\frac{\pi}{2}-y}{2}  \right]^{\frac{1}{2}} },\,\,\,\, \omega(x) \equiv \log \frac{\sin \frac{x+\frac{\pi}{2}}{2}}{\sin \frac{\frac{\pi}{2}-x}{2} }.
\end{equation}
Similarly, 
\begin{equation}
    H_{21}(x,y) = i\pi k(y,x)\frac{d}{d\omega(y)}\delta(\omega(y)-\omega(x)) = i\pi k(y,x)\left[ \frac{1}{\omega'(y)^2}\delta'(y-x)  - \frac{\omega''(y)}{\omega'(y)^3}\delta(x-y)  \right].
\end{equation}
By using
\begin{equation}
    k(x,x) = 2\omega'(x),\,\,\,\, \partial_x k(x,y)|_{y=x} = \omega''(x),
\end{equation}
we have
\begin{equation}\label{mh2}
    H_{12}(x,y) = -2\pi i \left[ \frac{1}{\omega'(x)}\delta'(x-y) + \frac{1}{2}\partial_x \left( \frac{1}{\omega'(x)} \right) \delta(x-y) \right],
\end{equation}
and
\begin{equation}\label{mh3}
    H_{21}(x,y) = 2\pi i\left[ \frac{1}{\omega'(y)}\delta'(y-x) + \frac{1}{2}\partial_y \left( \frac{1}{\omega'(y)} \right) \delta(y-x)  \right].
\end{equation}
Taking (\ref{mh2}) and (\ref{mh3}) back into (\ref{mh1}), we get
\begin{equation}
\begin{aligned}
    \mathcal{H} & =  2\pi \int_{0}^{\frac{\pi}{2}} dx \, \left\{\frac{1}{\omega'(x)}\frac{1}{2} \left[ i \partial_x \psi_1 (x)\psi_2 (x) -i\psi_1 (x)\partial_x\psi_2 (x)  \right] + \frac{i}{2}\partial_x \left(\psi_1 (x) \frac{1}{\omega'(x)}\psi_2(x)\right)\right\}\\
    & =  2\pi \int_{0}^{\frac{\pi}{2}}  dx \, \frac{1}{\omega'(x)} \frac{1}{2}\left[ i \psi_{-}(x)\partial_x \psi_- (x)  -i\psi_{+} (x)\partial_x\psi_{+} (x)  \right] + \pi i  \psi_1 (x) \frac{1}{\omega'(x)}\psi_2(x)|_{0}^{\frac{\pi}{2}}.
\end{aligned}
\end{equation}
Since $\frac{1}{\omega'(x)} = \cos x$, we get
\begin{equation}
    \mathcal{H} = 2\pi \int_{0}^{\frac{\pi}{2}}  dx\, \cos x\, T(x) - \pi i \psi_+ (0)\psi_- (0) ,
\end{equation}
where $T(x) = \frac{1}{2} \left[ i \psi_{-}(x)\partial_x \psi_- (x)  -i\psi_{+} (x)\partial_x\psi_{+} (x)  \right]$ is the energy density operator. By the boundary condition, $\psi_+ (0)\psi_- (0) = \psi_+ (0)^2 $ is a constant and can be dropped from the definition of the modular Hamiltonian. Thus we get
\begin{equation}
      \mathcal{H} = 2\pi \int_{0}^{\frac{\pi}{2}}  dx\, \cos x\, T(x) .
\end{equation}

\section{Negative energy for general $\Delta < 1/2$}\label{sec:massive}

In this section, we consider the negative energy generated by the boundary coupling for the cases of $\Delta < 1/2$. For simplicity, we consider massive free scalar field in AdS$_2$. We will calculate the energy perturbatively, by the same method as in \cite{Gao:2016bin}. 

After we finished this paper, we found a recent paper \cite{Bak:2019mjd} which involves similar calculation as in this appendix. In their case, the coupling is turned on as a constant at some intermediate time, and they study the backreaction on the dilaton field. Here we consider the case that we turn on the coupling adiabatically at $t=-\infty$, and we look at the bulk energy distribution of the matter field.

We consider a free massive scalar field $\chi$ in global AdS$_2$ spacetime, whose equation of motion is given by
\begin{equation}
     \left[ \sin ^ { 2 } \sigma \left( - \partial _ { t } ^ { 2 } + \partial _ { \sigma } ^ { 2 } \right) - h ( h - 1 ) \right] \chi = 0,
\end{equation}
where parameter $h$ is related to the mass of the scalar by $h(h-1) = m^2$. Here we changed notation from $\Delta$ to $h$ to be consistent with the literature. We are interested in the case that $h<1/2$. When we don't have the interaction between the two boundaries, we demand the boundary condition that the scalar field falls off at both boundaries as $\chi \sim \sin^h \sigma$. 

The normalized positive-frequency solutions are
\begin{equation}
      \chi _ { n } = \Gamma ( h ) 2 ^ { h - 1 } \sqrt { \frac { n ! } { \pi \Gamma ( n + 2 h ) } } e ^ { - i ( n + h ) t } ( \sin \sigma ) ^ { h } C _ { n } ^ { h } ( \cos \sigma ) , \quad n = 0,1 , \ldots.
\end{equation}
where $C_{n}^{h}$ is the Gegenbauer polynomial. The vacuum Wightman function is worked out by mode summation in \cite{Spradlin:1999bn} and \cite{Bardeen:1984hm}. The result is
\begin{equation}
      \langle \chi(x) \chi (y)\rangle = \frac { \Gamma ( h ) ^ { 2 } } { 4 \pi \Gamma ( 2 h ) }  \left( \frac{ 2} { d(x,y)} \right) ^ { h } F \left( h , h ; 2 h ; - \frac{ 2} { d(x,y)} \right) ,
\end{equation}
where
\begin{equation}
     d (x,y) = \frac { \cos \left( t _ { 1 } - t _ { 2 } \right) - \cos \left( \sigma _ { 1 } - \sigma _ { 2 } \right) } { \sin \sigma _ { 1 } \sin \sigma _ { 2 } }.
\end{equation}
The bulk to boundary propagator is given by
\begin{equation}
    K_{L}( t_1 - t_2 - i\epsilon,\sigma_1) \equiv  \left  \langle  \chi(t_1 ,\sigma_1) O_L \left(t_2 \right)\right \rangle =  \frac { \Gamma ( h ) ^ { 2 } } { 4 \pi \Gamma ( 2 h ) }  \left( \frac{ 2\sin \sigma_1 } {\cos (t_1 -t_2 - i\epsilon) - \cos \sigma_1 } \right) ^ { h },
\end{equation}
\begin{equation}
      K_{R}( t_1 - t_2 - i\epsilon,\sigma_1 ) \equiv  \left  \langle  \chi(t_1 ,\sigma_1) O_R \left(t_2 \right)\right \rangle =\frac { \Gamma ( h ) ^ { 2 } } { 4 \pi \Gamma ( 2 h ) }  \left( \frac{ 2\sin \sigma_1 } {\cos (t_1 -t_2 - i\epsilon) + \cos \sigma_1 } \right) ^ { h },
\end{equation}

To calculate the energy distribution at $t=0$, it's more convenient to work in Euclidean signature ($\tau = it$). The bulk to boundary propagator in Euclidean signature is given by
\begin{equation}
    \left\langle  \chi(\tau_1 ,\sigma_1) O_L \left(\tau_2 \right)\right \rangle =  \frac { \Gamma ( h ) ^ { 2 } } { 4 \pi \Gamma ( 2 h ) }  \left( \frac{ 2\sin \sigma_1 } {\cosh (\tau_1 - \tau_2) - \cos \sigma_1 } \right) ^ { h },
\end{equation}
\begin{equation}
    \left\langle  \chi(\tau_1 ,\sigma_1) O_R \left(\tau_2 \right)\right \rangle =  \frac { \Gamma ( h ) ^ { 2 } } { 4 \pi \Gamma ( 2 h ) }  \left( \frac{ 2\sin \sigma_1 } {\cosh (\tau_1 - \tau_2) + \cos \sigma_1 } \right) ^ { h }.
\end{equation}
Then we have
\begin{equation}\label{phi2}
\begin{aligned}
    \delta \langle \chi^2  (0,\sigma) \rangle & \approx -2g \int_{-\infty}^{\infty}d\tau \, \langle  \chi (0,\sigma) O_{L} (\tau)   \rangle \langle  \chi (0,\sigma) O_{R} (\tau)   \rangle \\
    & = -2g\left( \frac { \Gamma ( h ) ^ { 2 } } { 4 \pi \Gamma ( 2 h ) }  \right)^2   \left(2\sin \sigma \right)^{2h} \int_{-\infty}^{\infty} dt\, \frac{ 1 } { \left(\cosh^2 \tau - \cos^2 \sigma\right) ^ { h } } \\
    & = -2 g \left( \frac { \Gamma ( h ) ^ { 2 } } { 4 \pi \Gamma ( 2 h ) }  \right)^2  \left(2\sin \sigma \right)^{2h}  F\left(   h,h, h+\frac{1}{2}, \cos^2 \sigma \right) B\left( \frac{1}{2} , h \right)\\
    & = -\frac{g\Gamma (h)^3 }{2^{2h+1} \sqrt{\pi}\Gamma (\frac{1}{2} + h)^3} \sin^{2h} \sigma F\left(   h,h, h+\frac{1}{2}, \cos^2 \sigma \right).
\end{aligned}
\end{equation}
Similarly, we can get
\begin{equation}\label{dtphi2}
\begin{aligned}
    \delta \langle (\partial_t \chi)^2  \rangle =-  \delta \langle (\partial_\tau \chi)^2  \rangle \approx \frac{g2^{2h-3}h^2\Gamma (h)^4 B\left( \frac{3}{2},h  \right) }{ \pi^2 \Gamma (2h)^2 } \sin^{2h} \sigma  F\left( h+1,h,h+\frac{3}{2} ,\cos^2 \sigma \right),
\end{aligned}
\end{equation}

\begin{equation}\label{dxphi2}
\begin{aligned}
      \delta \langle (\partial_\sigma \chi)^2  \rangle 
   & \approx - \frac{gh^2 \Gamma (h)^3 }{2^{2h+1} \sqrt{\pi}\Gamma (\frac{1}{2} + h)^3} \sin^{2h-2} \sigma \cos^2 \sigma F\left(   h,h+1, h+\frac{1}{2}, \cos^2 \sigma \right)\\
   & \quad + \frac{g2^{2h-3} h^2 \Gamma (h)^4 B(\frac{1}{2},h+1) }{ \pi^2 \Gamma (2h)^2} \sin^{2h-2} \sigma  F\left(   h+1,h+1, h+\frac{3}{2}, \cos^2 \sigma \right).
\end{aligned}
\end{equation}

For $h\leq 1/2$, the naive definition of $T_{tt}$ will diverge as $\sin^{2h-2}\sigma$ near the boundary, and the energy is not integrable. Based on \cite{Breitenlohner:1982bm}, to remove the divergence of energy near the boundary of spacetime, one should add an improvement tern to the stress tensor:
\begin{equation}
    \tilde{T}_{\mu\nu} =   T _ { \mu \nu } + \beta \left( g_{\mu\nu} \nabla^2 - \nabla_\mu \nabla_\nu  + R_{\mu\nu} \right) \chi^2.
\end{equation}
In time independent cases, we have
\begin{equation}
    \tilde{T}_{tt} = T_{tt} +\beta \left(  \frac{1}{\sin^2 \sigma} - \frac{1}{\tan \sigma} \partial_{\sigma} - \partial_\sigma^2  \right)\chi^2.
\end{equation}
To cancel the divergence, we need to choose $\beta = h/(2(2h+1))$. With eqn. (\ref{phi2}), (\ref{dtphi2}) and (\ref{dxphi2}), we can calculate the energy induced by boundary interaction by
\begin{equation}
   \langle  \tilde{T}_{tt}  \rangle =  \frac{1}{2}\left(  \langle(\partial_t\chi)^2 \rangle+\langle(\partial_\sigma\chi)^2\rangle+\frac{h(h-1)\langle\chi^2\rangle}{\sin^2\sigma}\right)
   +\frac{h}{2(2h+1)} \left(  \frac{1}{\sin^2 \sigma} - \frac{1}{\tan \sigma} \partial_{\sigma} - \partial_\sigma^2  \right)\langle\chi^2 \rangle.
\end{equation}

\begin{figure}[t!]
 	\center
 	\includegraphics[width=12cm]{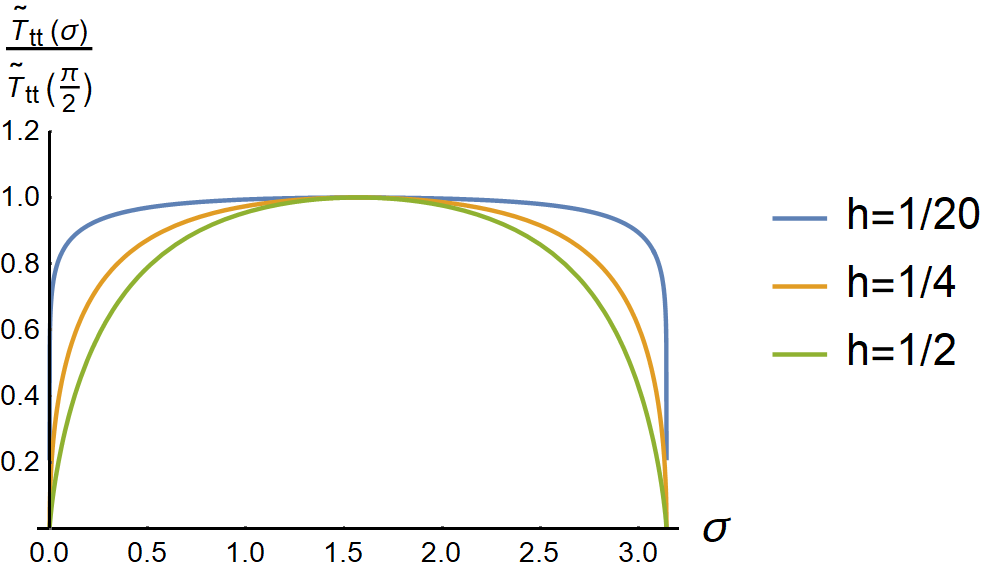}
 	\caption{Normalized energy distribution in the bulk for different $h\leq\frac{1}{2}$.} \label{energy1} 	
 \end{figure}

The final expression is

\begin{equation}
\begin{aligned}
    \langle  \tilde{T}_{tt}  \rangle \approx \frac{g 4^{-h} \Gamma (h+1)^3 }{\sqrt{\pi } (2 h+1)^2 \Gamma \left(h+\frac{1}{2}\right)^3}\sin ^{2 h}\sigma \, F\left(h,h+1;h+\frac{3}{2};\cos
   ^2\sigma\right).
\end{aligned}
\end{equation}

We find that $\tilde{T}_{tt}$ scales as $\sin^{2h} \sigma$ near the boundary, and is thus integrable. For $h=0$, the energy is a constant in the bulk, which is consistent as being a conformal field. For $h=1/2$, the energy scales as $\sin \sigma$ near the boundary (see fig. \ref{energy1}). We see that for $h\leq 1/2$, the energy is more concentrated in the center of the bulk than near the boundary. This is consistent with the intuition for relevant perturbation in AdS/CFT.

\section{Solving the dilaton profile for $t>0$}\label{sec:solvedilaton}

\begin{figure}[t!]
    \centering
    \includegraphics[width=6cm]{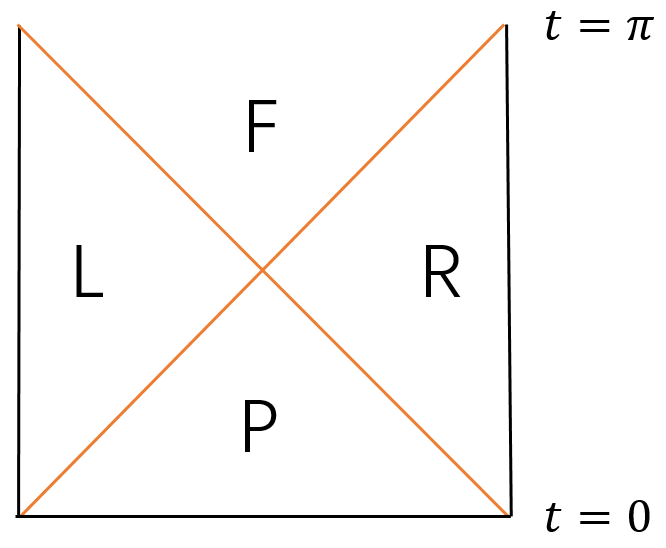}
    \caption{The two shock waves (orange lines) divide the bulk into four regions, which we denote as L (left), R (right), F (future), P (past).}
    \label{fig:appendixdilaton}
\end{figure}

In this appendix, we present the detail of solving the dilaton profile for the case of a sudden quench in section. \ref{evolution}. As discussed in sec. \ref{evolution}, for $0<t<\pi$, the energy distribution in the bulk is:
\begin{equation}
    T_{++}^{M} (x^+)/N = - \frac{\epsilon}{8\pi} + \frac{\epsilon}{8} \delta (x^+ - \pi + \delta) ,\,\,\,\,   T_{--}^{M} (x^-)/N = - \frac{\epsilon}{8\pi} + \frac{\epsilon}{8} \delta (x^- + \delta) .
\end{equation}
As an example, let's look at how to derive the dilaton field profile in region L in fig. \ref{fig:appendixdilaton}.

Since in region P and L, we both have constant negative energy in the bulk, the solution in L can be written as
\begin{equation}
    \phi_{L} = \phi_{P} + \varphi_L,
\end{equation}
where $\varphi_L$ has support only in regions L and F. From 
\begin{equation}
    \partial_{-} \left(  \sin^2 \sigma \partial_- \varphi_L   \right) = -N\frac{\epsilon}{8} \delta (x^- + \delta) \sin^2 \sigma,
\end{equation}
we get
\begin{equation}
    \partial_{-}\varphi_L = -N\frac{\epsilon}{8}  \frac{ \sin^2 \frac{x^+ + \delta}{2}}{\sin^2 \frac{x^+ - x^-}{2}},
\end{equation}
Integrate it, we get
\begin{equation}
    \varphi_L (x^+ , x^-) = - N\frac{\epsilon}{4} \sin^2 \frac{x^+ + \delta}{2} \left( \frac{1}{\tan\frac{x^+ - x^-}{2} } -   \frac{1}{\tan\frac{x^+  + \delta}{2} }   \right).
\end{equation}
Thus
\begin{equation}
\begin{aligned}
   \phi_{L}/N & =  \frac{\epsilon}{4\pi} \left( 1+  \frac{\frac{\pi}{2} - \frac{x^+ - x^-}{2}}{\tan\frac{x^+ - x^-}{2}} \right) - \frac{\epsilon}{4} \sin^2 \frac{x^+ + \delta}{2} \left( \frac{1}{\tan\frac{x^+ - x^-}{2} } -   \frac{1}{\tan\frac{x^+  + \delta}{2} }   \right)\\
   & = \frac{\epsilon}{4\pi} \left( 1+  \frac{\frac{\pi}{2} - \sigma}{\tan\sigma} \right) - \frac{\epsilon}{4} \sin^2 \left( \frac{\sigma + t + \delta}{2}  \right) \left( \frac{1}{\tan\sigma } -   \frac{1}{\tan\frac{\sigma + t + \delta}{2}  }   \right)\\
   & = \frac{\epsilon}{4\pi} \left( 1+  \frac{\frac{\pi}{2} - \sigma}{\tan\sigma} \right) - \frac{\epsilon}{4}  \frac{\sin^2\frac{\sigma + t + \delta}{2} \cos \sigma - \sin  \sigma \sin\frac{\sigma + t + \delta}{2}\cos \frac{\sigma + t + \delta}{2}}{\sin \sigma}\\
   & =  \frac{\epsilon}{4\pi} \left( 1+  \frac{\frac{\pi}{2} - \sigma}{\tan\sigma} - \frac{\pi}{2}\frac{\cos \sigma - \cos (\sigma + t + \delta) \cos \sigma - \sin  \sigma \sin(\sigma + t + \delta) }{\sin \sigma}  \right)\\
   & =  \frac{\epsilon}{4\pi} \left( 1 -   \frac{ \sigma}{\tan\sigma} + \frac{\pi}{2}\frac{\cos (t +\delta ) }{\sin \sigma} \right).
\end{aligned}
\end{equation}
One can check that this solution satisfies the $T_{++}^{M}$ and $T_{+-}^{M}$ equations. The solution in region R can be derived in the same way:
\begin{equation}
    \phi_R = \phi_P + \varphi_R,
\end{equation}
where $\varphi_R$ is simply a reflection of $\varphi_L$ upon $\sigma = \pi/2$. Finally, the solution in region F is:
\begin{equation}
    \phi_F = \phi_P + \varphi_L + \varphi_{R}.
\end{equation}

 \end{appendices}

\bibliographystyle{JHEP}
\nocite{*}
\bibliography{cite}

\end{document}